%% file: desi_paper_main_PASA.tex
\documentclass[
  journal=pasa,
  manuscript=article-type,
  year=2020,
  volume=37,
]{cup-journal}

\usepackage{booktabs}
\usepackage{mathptmx}
\usepackage{inconsolata}
\usepackage[T1]{fontenc}
\usepackage{comment}
\usepackage{nicefrac}
\usepackage{caption}
\usepackage{graphicx}	
\usepackage{xcolor}
\usepackage{float}
\usepackage{setspace}	
\usepackage{microtype}
\usepackage[super]{nth}
\usepackage{siunitx}
\usepackage{numprint}
\usepackage{amsmath, mathtools, amsfonts}
\usepackage{subcaption}
\usepackage{multirow}
\usepackage{array}
\usepackage{longtable}
\usepackage{pdflscape}
\usepackage{hyperref}
\usepackage{siunitx}
\usepackage{subcaption}
\usepackage{longtable}
\usepackage{placeins}
\usepackage{caption}
\usepackage{amssymb}

\DeclareRobustCommand{\lapprox}{%
  \ensuremath{%
    \mathrel{%
      \ooalign{%
        $<$\cr
        \hidewidth\lower0.3ex\hbox{$\scriptscriptstyle\sim$}\hidewidth\cr
      }%
    }%
  }%
}
\DeclareRobustCommand{\gapprox}{%
  \ensuremath{%
    \mathrel{%
      \ooalign{%
        $>$\cr
        \hidewidth\lower0.3ex\hbox{$\scriptscriptstyle\sim$}\hidewidth\cr
      }%
    }%
  }%
}

\title{Photometric Redshift Predictions with a Neural Network for DESI Quasars}

\author{J. P. Moss}
\affiliation{School of Chemical and Physical Sciences, Victoria University of Wellington, Wellington, New Zealand}
\email[J. P. Moss]{mossji@staff.vuw.ac.nz}

\author{S. J. Curran}
\affiliation{School of Chemical and Physical Sciences, Victoria University of Wellington, Wellington, New Zealand}

\author{Y. C. Perrott}
\affiliation{School of Chemical and Physical Sciences, Victoria University of Wellington, Wellington, New Zealand}


\begin{document}

\begin{abstract}
Accurate redshift measurements are essential for studying the evolution of quasi-stellar objects (QSOs) and their role in cosmic structure formation. While spectroscopic redshifts provide high precision, they are impractical for the vast number of sources detected in large-scale surveys. Photometric redshifts, derived from broadband fluxes, offer an efficient alternative, particularly when combined with machine learning techniques. In this work, we develop and evaluate a neural network  model for predicting the redshifts of QSOs in the Dark Energy Spectroscopic Instrument (DESI) Early Data Release spectroscopic catalogue, using photometry from DESI, the Widefield Infrared Survey Explorer (WISE) and the Galactic Evolution Explorer (GALEX). We compare the performance of the neural network model against a k-Nearest Neighbours approach, these being the most accurate and least resource-intensive of the methods trialled herein, optimising model parameters and assessing accuracy with standard statistical metrics. Our results show that incorporating ultraviolet photometry from GALEX improves photometric redshift estimates, reducing scatter and catastrophic outliers compared to models trained only on near infrared and optical bands. The neural network achieves a correlation coefficient with spectroscopic redshift of $0.9187$ with normalised median absolute deviation of $0.197$, representing a significant improvement over other methods. Our work combines DESI, WISE and GALEX measurements, providing robust predictions which address the difficulties in predicting photometric redshift of QSOs over a large redshift range. 
\end{abstract}

\input{desi_paper_v3}

\clearpage
\bibliographystyle{pasa-mnras}
\bibliography{cleaned_master_bibliography_from_bbl} 
\end{document}

%% file: desi_paper_v3.tex

\section{Introduction}
\subsection{Background}
    Accurate redshift measurement is crucial for understanding the cosmological and physical properties of active galactic nuclei (AGN), particularly quasars, which are the most luminous and distant objects in the Universe. Although spectroscopic measurements provide precise redshifts, they are resource-intensive and impractical for the vast number of sources that upcoming surveys, such as those with the Square Kilometre Array (SKA; \citealt{carilli2004, schilizzi2004, norris2011emu}), will detect. This highlights the need for efficient photometric redshift estimation methods that can handle large datasets. Photometric redshifts, hereafter referred to as $z_\mathrm{phot}$, derived from broadband flux measurements in the ultraviolet (UV), optical and near infrared bands (NIR), offer a complementary approach, enabling the rapid analysis of quasars and other celestial objects. Using machine learning techniques trained on large datasets, these methods can provide reliable redshift estimates, paving the way for timely data processing and analysis in the SKA and other large-scale surveys.


    The identification and classification of AGN such as quasars and quasi-stellar objects (QSOs) requires the detection of spectral lines. However, even with the development of multiobject spectrometers \citep{wolf2018}, spectroscopy is a time- and resource-intensive process \citep{popowicz2017}. Acquiring high-quality spectra is a demanding task due to the necessity for high signal-to-noise ratios and spectral resolutions. Achieving resolutions of $R=\num{300000}$ or higher is essential for fully resolving line shapes and accurately interpreting wavelength shifts in QSO spectra \citep{dravins2010}. However, several challenges can complicate the interpretation of these spectra, such as incomplete spectroscopic data \citep{connolly1999}, the absence of suitable lines, overlapping lines from different sources, and imprecise laboratory wavelengths. Additionally, obtaining such high-fidelity spectra often requires long integration times. The development of efficient spectrometers with resolutions approaching $R=\num{1000000}$ for future large telescopes remains a significant challenge in advancing our understanding of QSO spectra \citep{dravins2010}, and the spectroscopic redshifts ($z_\mathrm{spec}$) derived from them. However, higher spectral resolution also comes at the cost of increased noise, necessitating even longer exposure times to maintain adequate signal-to-noise-ratios.

    Photometric redshift estimates based on broadband photometry or template-fitting (see, for example, \citealt{Ball2007, Carliles2007, bovy2012, moss2021_ML, zhou2021, moss2022_milliquas} and references therein) from the photometry of known objects provide a valuable alternative but are often plagued by inherent uncertainties. Once trained on a sufficiently large dataset, modern machine learning (ML) models allow the analysis of large amounts of data, from which they can estimate the redshift from the photometric measurements taken by the survey. This enables us to predict the redshift of a large number of sources without relying on a spectrum for each object. As modern surveys collect vast amounts of data, the ability to rapidly estimate redshifts using ML models becomes crucial for timely data processing and release. By validating and refining photometric redshift methods against spectroscopic data, we can improve the accuracy of these models, extending their applicability to future surveys and legacy datasets. Such methods also hold potential for identifying high-redshift objects in upcoming surveys, where spectroscopic follow-up may be limited or delayed.
    Developing robust methods and pipelines for estimating redshifts from photometric measurements is essential to maximise the scientific potential of upcoming surveys.



    In this paper, we present a neural network capable of predicting QSO redshifts in the Dark Energy Spectroscopic Instrument (DESI; \citealt{dey2019, desicollaboration2022, desicollaboration2023, chaussidon2023, alexander2023}) dataset with an accuracy of $\sim81\%$, which increases to $\sim92\%$ with the inclusion of photometry from Galaxy Evolution Explorer (GALEX; \citealt{Martin2005, gildepaz2007}).

\section{Data and Methods}\label{sec:Data}
    \subsection{DESI}\label{sec:DESI_data}
        DESI surveys the sky in the $-34^{\circ} < \delta \leq 90^{\circ}$ declination range using the $g$, $r$, and $z$ bands. Data Release 9 (DR9 \citealt{schlegel2021}) includes images and photometric measurements of 2.85 million sources. The photometry of these sources from DESI is complemented by forced photometry in the $W1$ and $W2$ bands from unWISE coadded images, derived from Wide-field Infrared Explorer (WISE) mission \citep{wright2010}; that is, flux measurements were extracted at the positions of sources detected in the optical bands, even if the sources were too faint to be independently detected in WISE infrared (IR) images.
        
\begin{figure}
            \centering
            \includegraphics[width=1\linewidth]{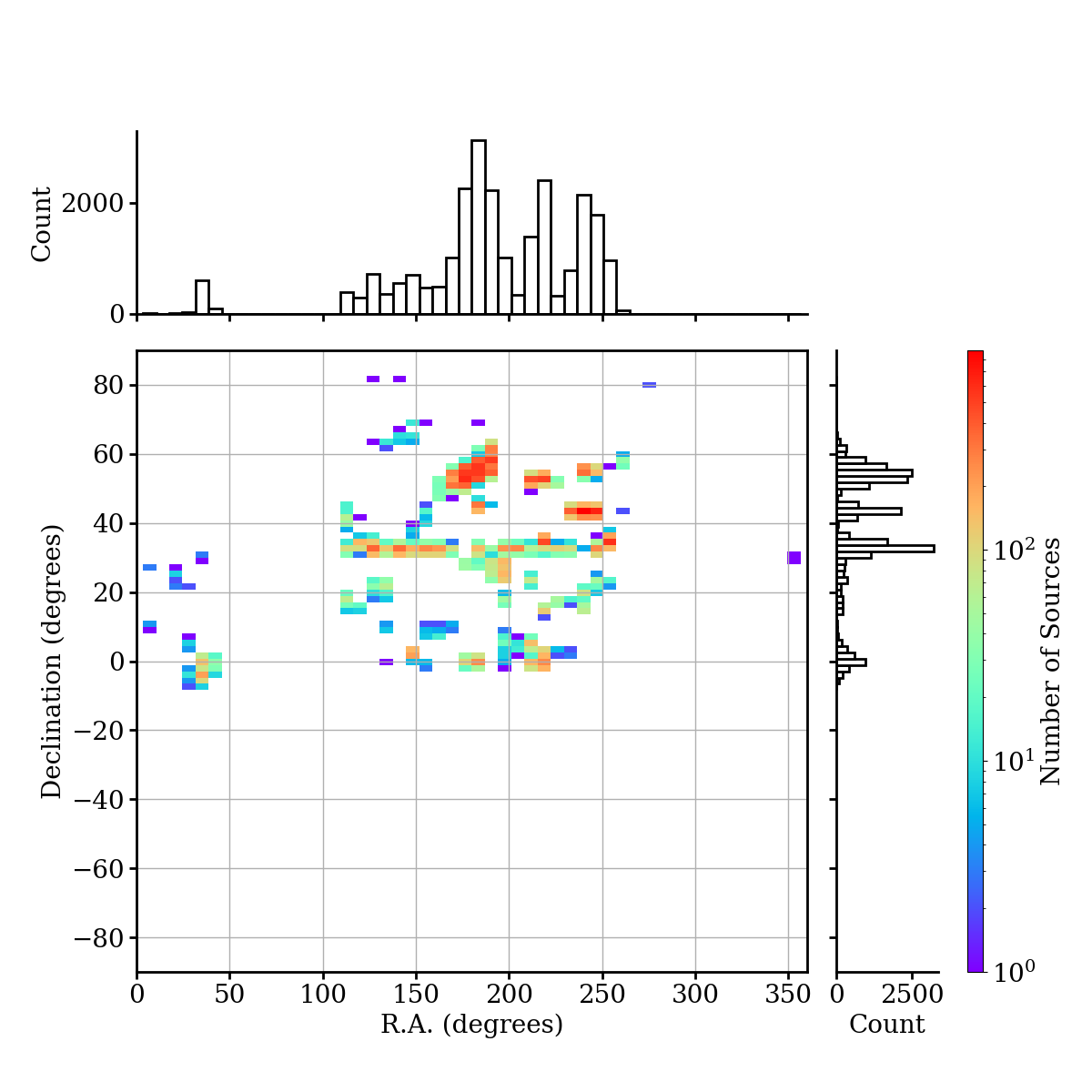}
            \includegraphics[width=1\linewidth]{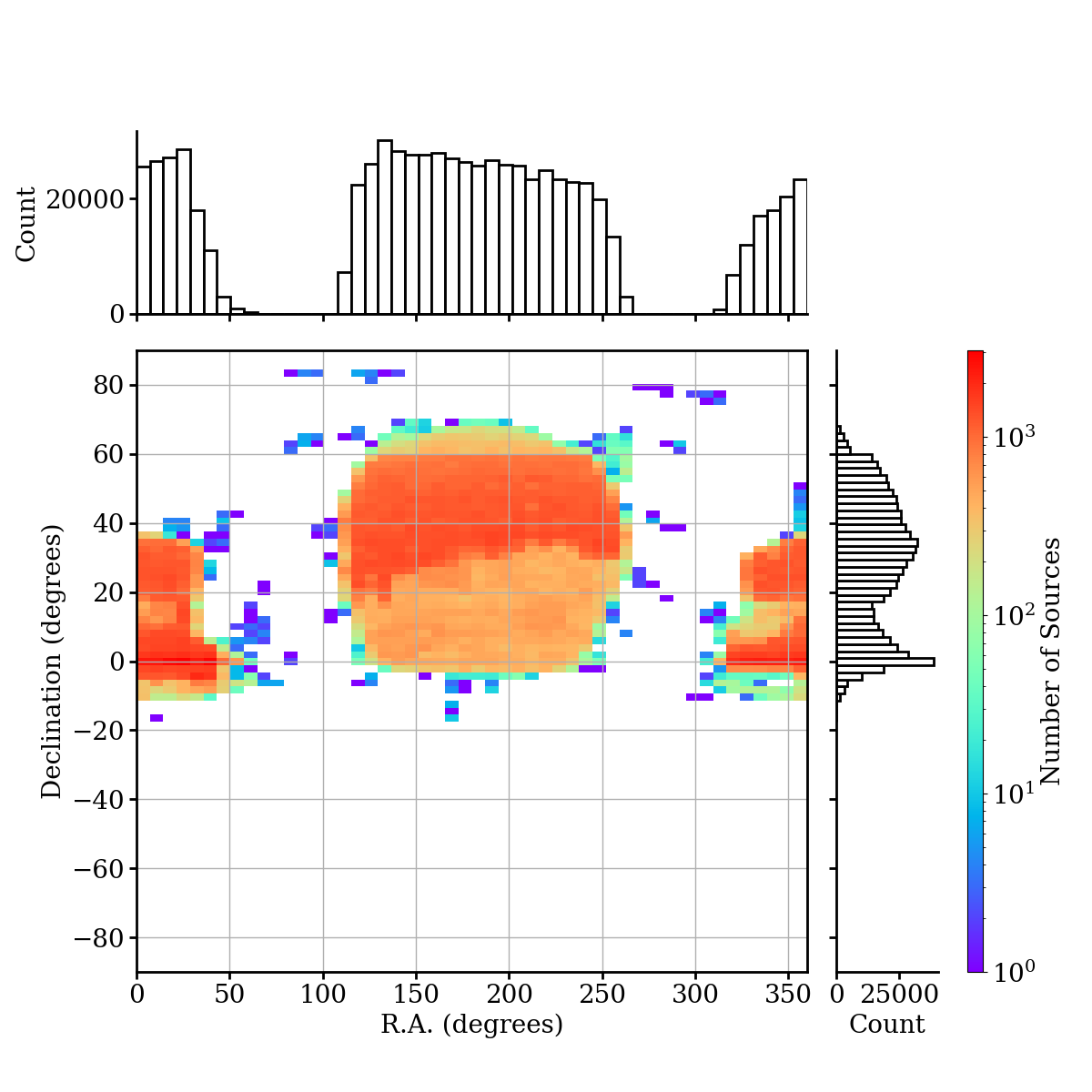}
            \caption{The sky distributions of the DESI EDR QSO spectroscopic catalogue (top, from \citealt{desicollaboration2023}) and SDSS (bottom, from \citealt{lyke2020}) samples. The histograms show the number of sources in right ascension and declination.}
            \label{fig:sky_dist}
        \end{figure}

        This imaging data serves as the basis for target selection in the DESI spectroscopic survey. The Early Data Release (EDR) QSO catalogue\footnote{\texttt{QSO\_cat\_fuji\_healpix\_only\_qso\_targets.fits} available at \url{https://data.desi.lbl.gov/public/edr/vac/edr/qso/v1.0/}}, for which the sky distribution is shown in Figure~\ref{fig:sky_dist}, contains \num{87318} sources spectroscopically identified as QSOs using DESI's Redrock (RR) template-fitting algorithm and the QuasarNET (QN) deep-learning classifier. In this paper, the term `DESI dataset' encompasses both the imaging data from the Legacy Imaging Surveys' DR9, which includes $g$, $r$, and $z$ fluxes, and the spectroscopic data from the EDR QSO catalogue, which provides redshift information for a subset of these sources. Both include $W1$ and $W2$ fluxes from unWISE coadded images, derived using forced photometry as described above. Distributions of the fluxes in the DESI dataset are shown in Figure~\ref{fig:DESI_flux_dist}.

    \subsection{DESI x SDSS}\label{sec:DESIxSDSS_data}
        To assess the impact of missing photometric bands on redshift prediction performance, the DESI dataset was crossmatched with the Sloan Digital Sky Survey (SDSS) Data Release 16 QSO catalogue (DR16Q; \citealt{lyke2020}). The DESI component of the dataset, described in Section~\ref{sec:DESI_data}, provided photometry in the $g$, $r$, and $z$ bands, along with infrared fluxes from WISE ($W1$ and $W2$).
        
        SDSS contains optical magnitudes ($u$, $g$, $r$, $i$, $z$), where $g$, $r$, and $z$ overlap with DESI, and forced-photometry UV measurements in the far- and near-ultraviolet ($FUV$ and $NUV$) wavebands from GALEX for \num{750414} QSO sources. This crossmatch allows us to evaluate whether excluding the SDSS $u$ and $i$ bands, as well as the GALEX $FUV$ and $NUV$ bands, leads to a significant degradation in photometric redshift predictions.

\begin{figure}
            \centering
            \includegraphics[width=1.1\linewidth]{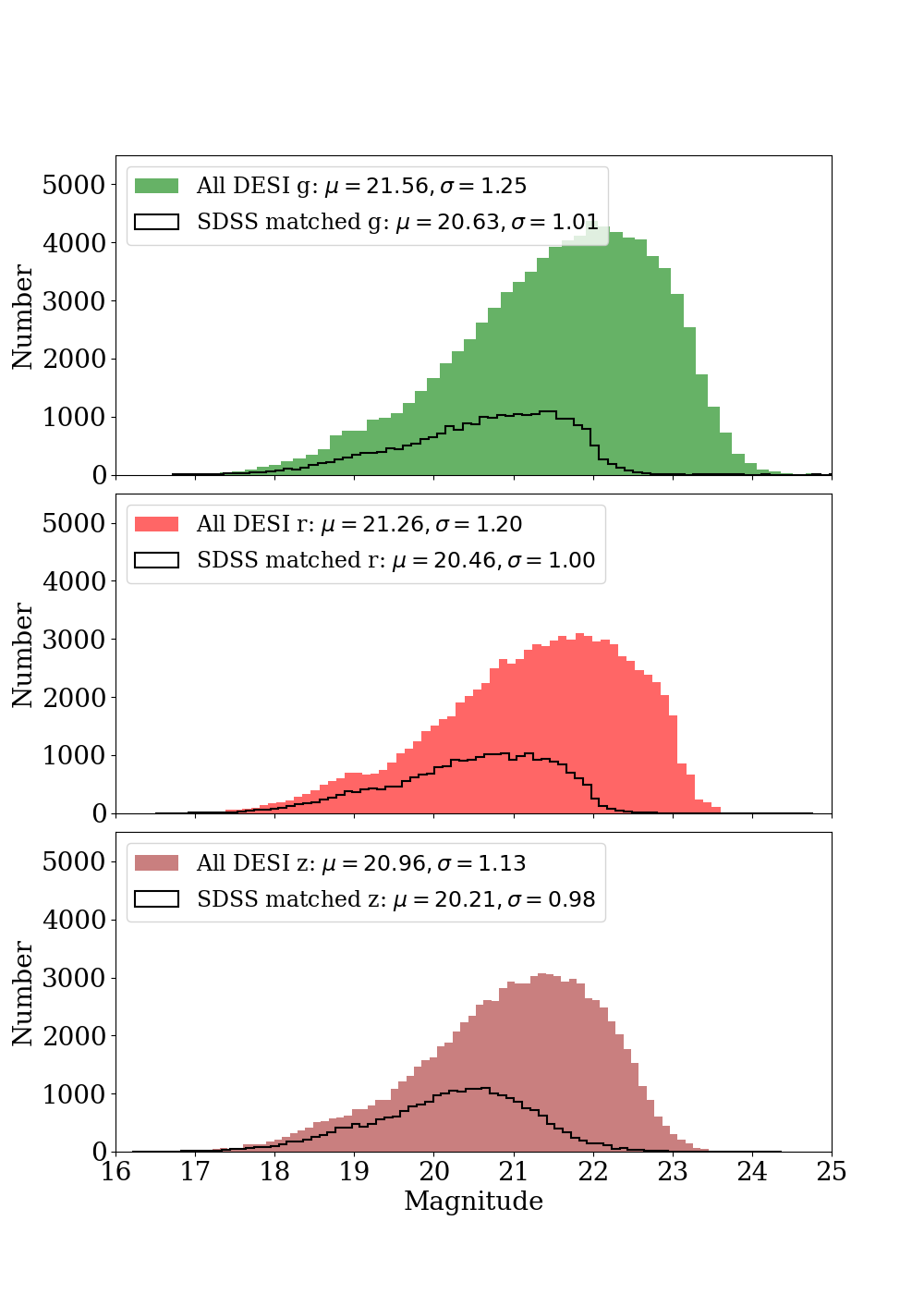}
            \caption{The distribution of magnitudes for the DESI sample as a whole compared to those with a match in SDSS (see Section~\ref{sec:SDSS}). The legend in each panel shows the mean magnitude and the standard deviation. While DESI fluxes are used directly for model training, the comparison in this figure is made in magnitude space to match the SDSS format.}
            \label{fig:DESI_flux_dist}
        \end{figure}

\begin{figure}
            \centering
            \includegraphics[width=1\linewidth]{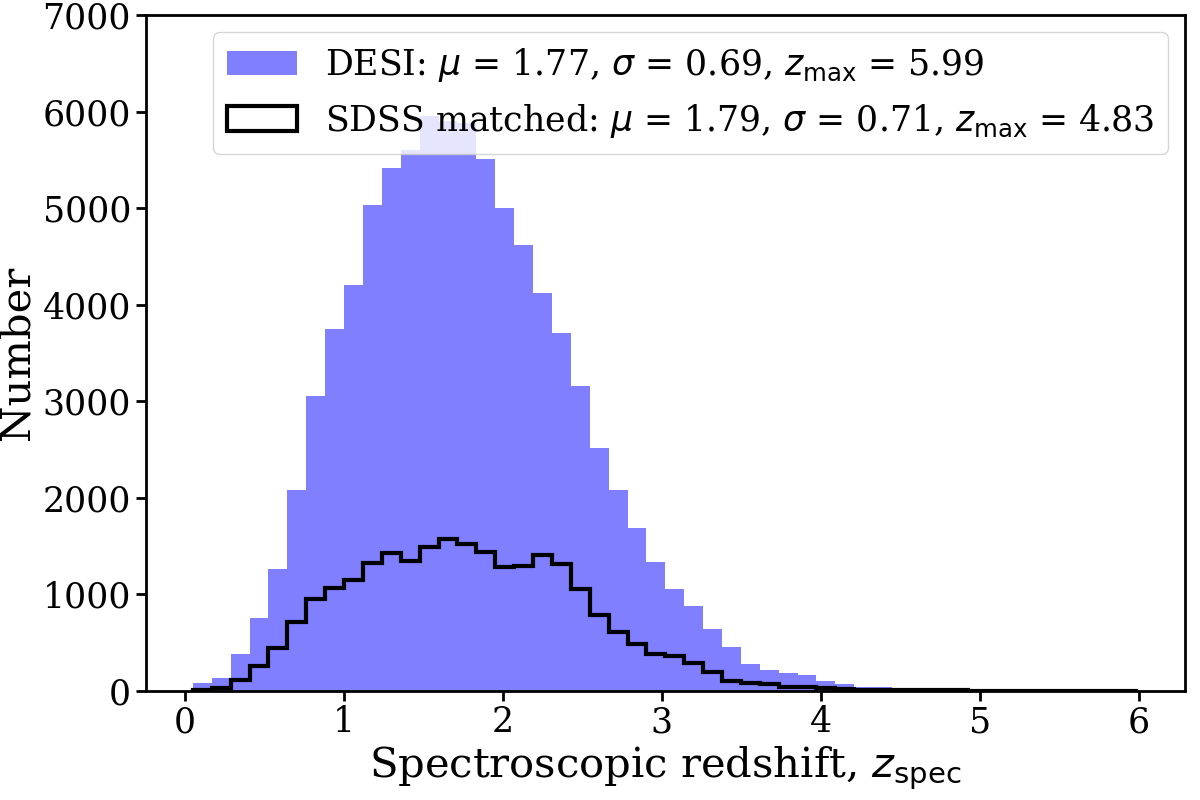}
            \caption{The distribution of redshifts for the full DESI sample and the SDSS-matched subset. The legend in each panel shows the mean redshift, standard deviation and maximum redshift.}
            \label{fig:z_dist_DESI}
        \end{figure}

        SDSS imaging has a median seeing of 1.32 arcseconds in the $r$-band\footnote{\url{https://www.sdss4.org/dr17/imaging/other_info}}, while DESI’s spatial resolution is limited by its 1.5 arcsecond fibre diameter\footnote{\url{https://noirlab.edu/public/media/archives/brochures/pdf/brochure007.pdf}}. A positional tolerance of 1 arcsecond was chosen for crossmatching to ensure high-confidence associations, returning 24 616 secure matches (only $3.2\%$ of the DESI EDR QSO sample) which we refer to as DxS. This modest fraction arises because SDSS imaging is both shallower ($z \lapprox 22\,\mathrm{mag}$) and covers a slightly different footprint than the DESI EDR, so many of the fainter or uniquely targeted DESI QSOs simply have no SDSS counterpart. Figure \ref{fig:sky_dist} illustrates the limited overlap, while Figures~\ref{fig:DESI_flux_dist} and \ref{fig:z_dist_DESI} show that, despite the low match fraction and the difference in magnitude distributions, the two redshift distributions are broadly similar in shape and spread, diverging mainly at the highest redshifts where SDSS drops out.
        
    \subsection{GALEX}\label{sec:GALEX_data}
\begin{figure}
            \centering
            \includegraphics[width=1\linewidth]{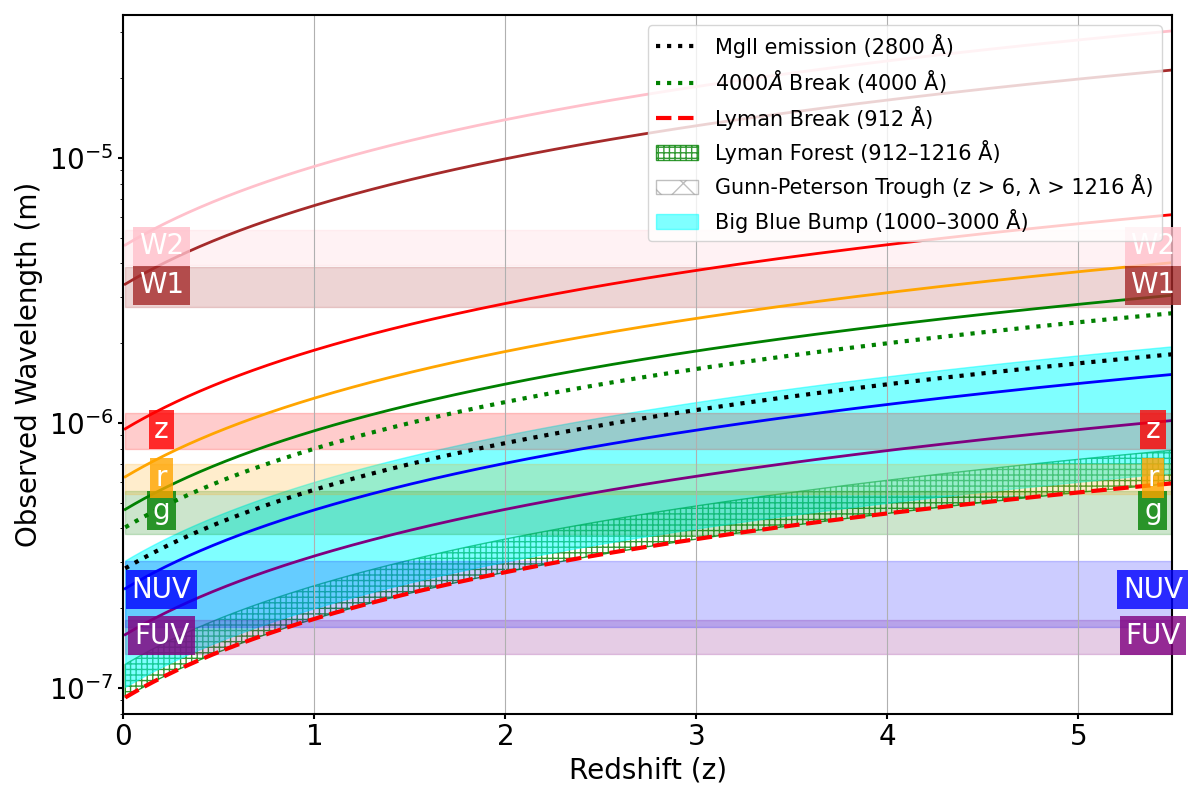}
            \caption{The variation of source-frame wavelength with redshift for $g$, $r$, $z$, $W1$, $W2$ and GALEX bands. The coloured horizontal bands show the `windows' provided by the filter ranges at the observed-frame wavelengths and the curves show those rest-frame wavelengths as a function of source redshift. The dashed red line represents the Lyman break ($\lambda = 1.216\times 10^{-7}$ m), the green square hashed region represents the Lyman forest and the shaded cyan region shows the Big Blue Bump. The black dotted line represents the Mg\,{\sc ii} emission line ($\lambda = 2.8\times 10^{-7}$ m) and the green dotted line shows the 4000 \r{AA} break. Labels identify the bands and lines. For example, for a source at a redshift of $z=5$, the $FUV$ line ($\lambda_\text{rest}=1.575\times10^{-7}$ m) has been shifted into the $z$-band.}
            \label{fig:variation_of_wavelength_with_redshift}
        \end{figure}

        GALEX provides UV photometry in two bands: $FUV$ and $NUV$. These bands are particularly valuable for studies of QSOs, as they probe rest-frame UV features that shift into optical wavelengths at high redshifts, as shown in Figure~\ref{fig:variation_of_wavelength_with_redshift}. For this study, GALEX measurements were obtained from the SDSS DR16Q catalogue, including forced-photometry UV fluxes for a subset of SDSS QSOs. We also attempted to crossmatch the DESI dataset directly with the standalone GALEX QSO catalogue \citep{atlee2007}, which contains \num{36120} sources with NUV and /or FUV fluxes. However, this yielded only \num{504} matches, significantly reducing the sample size, making it unusable as a training set for machine learning. As a result, we chose to use the GALEX UV fluxes available via the SDSS DR16Q catalogue, which provided broader coverage while maintaining consistency with the rest of the dataset. While a direct cross-match between the DESI quasar sample and GALEX yields only 504 high-confidence matches, the broader DxS dataset contains 24,616 sources with GALEX fluxes. This is because the GALEX photometry for DxS is sourced from the GALEX–SDSS cross-match catalogue, which includes both detections and forced photometry at SDSS positions. As such, many sources in the DxS dataset have GALEX fluxes even when no significant UV detection was made by SDSS. These fluxes often have large uncertainties or are upper-limit estimates. Although the inclusion of GALEX bands improves neural network predictions on average, the comparatively poor quality of the UV measurements likely contributes to the more modest or statistically insignificant improvements observed for the kNN model. The results highlight the importance of UV coverage for constraining redshifts, but also the need for caution when interpreting results based on noisy or uncertain fluxes.
        
        Previous studies have shown that a wide range of wavelengths is desirable for estimating photometric redshift due to the shifting of rest-frame wavelengths through filter bands \citep{Brescia2021, duncan2022}. The wide wavelength coverage of $W1$ - $FUV$ allows us to trace rest-frame features across redshift.

    
    \subsection{Preprocessing}\label{sec:Preprocessing}
        While we do have access to the redshift quality flags summarised in Table~\ref{tab:filtering} during training, we chose not to apply these filtering constraints so that the model remains applicable to real-world data where such quality indicators may be absent. We tested the effect of applying these filters in preliminary runs and found no significant difference in results, so the constraints were not used in the final training set.
        
        Before incorporating the photometric data into the training features, SDSS magnitudes and DESI fluxes were corrected for Galactic extinction using the \cite{schlafly2011} dust maps, and the data were standardised by the mean and standard deviation to ensure that the fluxes and magnitudes were comparable, enabling consistent input features for the models. 
        \begin{table}
            \centering
            \caption{Filtering criteria for the DxS sample and the number of sources affected by each filter. Names are as follows \citep{desicollaboration2023}: \texttt{ZERR} is the uncertainty in the spectroscopic redshift; \texttt{ZWARN} is a bitmask indicating if there are any known problems with the data or the spectroscopic fit; \texttt{SPECTYPE} is the spectral classification, which could be \texttt{STAR}, \texttt{GALAXY} or \texttt{QSO}.}
            \begin{tabular}{l r}
                \hline
                Filter criterion & Number of sources \\
                \hline
                \texttt{ZERR > 0.001} & 25 \\
                \texttt{ZWARN != 0} & 277 \\
                \texttt{SPECTYPE != `QSO'} & 1447 \\
                \hline
            \end{tabular}
            \label{tab:filtering}
        \end{table}


        Many ML models use the $u-g$, $g-r$, $r-i$ and $i-z$ colours of sources in SDSS to train and validate an ML model (for example, \citealt{Carliles2007, hoyle2015, Pasquet2019, li2022}). Throughout this study, we use raw fluxes and magnitudes as input features for the ML models, rather than colour indices, since our previous studies have shown that the raw measurements yield comparable results \citep{moss2021_ML, curran2022_incomplete}. This approach avoids introducing additional correlations between features, and retains the direct photometric measurements, ensuring that the models are trained on the most fundamental observational data. Across this work, the choice of using fluxes or magnitudes was guided by dataset conventions and practical considerations. For the DESI and DxS datasets, fluxes are used directly as provided in the Legacy Surveys DR9 catalogue. In contrast, SDSS photometry is typically provided in asinh magnitudes, which we retain. While this means the units differ between datasets, all fluxes and magnitudes are standardised prior to model training, minimising the impact of unit differences and ensuring that fluxes and magnitudes are directly comparable. This preprocessing step prevents unit-driven bias and ensures the models remain sensitive to relative patterns rather than absolute scales. No conversion between fluxes and magnitudes was therefore necessary. This choice also preserves the numerical properties of low-S/N sources, particularly in the case of asinh magnitudes and forced photometry. The choice to work with raw fluxes and magnitudes aligns with the goal of maintaining flexibility and generality for application across various surveys and datasets. 

    \subsection{Model Evaluation and Metrics}

    To evaluate model performance we use the following statistics:
\begin{itemize}
    \item \textit{Correlation coefficient} ($r$) of the least-squares linear fit between the predicted redshifts ($z_{\text{phot}}$) and the spectroscopic redshifts ($z_{\text{spec}}$). Perfect agreement would yield $r = 1$.

    \item \textit{Explained variance} (EV).  
          Define the residuals
          \[
              \Delta z \equiv z_{\text{spec}} - z_{\text{phot}} .
          \]
          EV compares the spread of these residuals with the intrinsic spread of the true redshifts:
          \[
              \mathrm{EV}
              = 1 - \frac{\sigma^{2}_{\Delta z}}{\sigma^{2}_{z_{\text{spec}}}},
          \]
          where the variance is
          \[
              \sigma^{2}_{x} = \frac{\sum_{i=1}^{N}\left(\Delta z\right)^{2}}{z_{\text{spec}}}.
          \]
          An ideal model has $\sigma^{2}_{\Delta z}\!\to\!0$ and hence $\mathrm{EV}\!\to\!1$.

    \item \textit{Normalised median absolute deviation} (NMAD), a robust scatter estimator insensitive to outliers:
          \[
              \sigma_{\mathrm{NMAD}}
              = 1.4826 \times \mathrm{median}\!\left(
                 \frac{|\Delta z|}{1 + z_{\text{spec}}}
                \right).
          \]

    \item \textit{Maximum error}
          \[
              e_{\max} = \max\!\bigl(|\Delta z|\bigr).
          \]

    \item \textit{Mean absolute error} (MAE)
          \[
              \mathrm{MAE}
              = \frac{1}{N}\sum_{i=1}^{N}|\Delta z_{i}|.
          \]
\end{itemize}

    Each model was trained on 80\% of the dataset, with 20\% reserved for testing. Models were trained on selected features, and the models were run 100 times. A summary of the features in each dataset is given in Table~\ref{tab:missing_data_summary}. Each run used its own train-test split so that the data were shuffled each time. The training utilised K-fold cross-validation, using the above metrics to quantify predictive performance. The average performance of each model across the 100 runs was computed, and the results inspected visually by comparing the $z_\mathrm{phot}$ generated by each model to the $z_\mathrm{spec}$ in the datasets, including their residuals, defined by $\Delta z = z_\mathrm{phot} - z_\mathrm{spec} $ with standard deviation $\sigma_{\Delta z}$.

\begin{table}
    
    \caption{Missing values by photometric band for each dataset. All datasets use extinction-corrected magnitudes or fluxes where applicable. GALEX fluxes include both direct detections and forced photometry. Only 504 DESI sources were matched to GALEX with reliable UV fluxes.}
    \centering
    \begin{tabular}{llllr}
        \hline
        \textbf{Dataset} & \textbf{Band} & \textbf{Type} & \textbf{Missing} & \textbf{\% Missing} \\
        \hline
        \textbf{SDSS (\num{750414})} & $u$     & mag  & 1,014 & 0.14 \\
                       & $g$     & mag  & 1,015 & 0.14 \\
                       & $r$     & mag  & 1,011 & 0.13 \\
                       & $i$     & mag  & 1,011 & 0.13 \\
                       & $z$     & mag  & 1,011 & 0.13 \\
                       & $FUV$, $NUV$ & flux & 0     & 0.00 \\
        \hline
        \textbf{DxS (\num{24616})}   & $g$, $r$, $z$     & flux & 0   & 0.00 \\
                       & $W1$, $W2$        & flux & 0   & 0.00 \\
                       
        \hline
        \textbf{DESI (\num{87318})} & $g$, $r$, $z$     & flux & 0   & 0.00 \\
                      & $u$, $g$, $r$, $i$, $z$ & mag  & 27  & 0.11 \\
                      & $W1$, $W2$        & flux & 0   & 0.00 \\
                      & $FUV$, $NUV$      & flux & 504 & 0.58 \\
        \hline
    \end{tabular}
    \label{tab:missing_data_summary}
\end{table}


\subsection{Neural Network}\label{sec:NN}
    In preliminary trials, several machine learning algorithms were evaluated, including \textit{Random Forest} (RF), \textit{Ridge Regression} (RR), \textit{Support Vector Machines} (SVM), \textit{ElasticNet} (EN), \textit{k-Nearest Neighbours} (kNN), and a \textit{Neural Network} (NN). The RF and SVM models yielded reasonably accurate predictions but were found to be prohibitively slow to train and test across the large dataset and multiple iterations required for this study. In contrast, RR and EN were computationally efficient but underperformed in terms of predictive accuracy, particularly at higher redshifts and in regions with sparse training data. Consequently, this paper focuses on the methodologies and results of the NN and kNN approaches. For the neural network, architectures with 1 to 6 hidden layers and 50 to 300 neurons per layer were explored, using ReLU activations, the Adam and AdamW optimisers, and mean squared error loss. The best configuration was selected based on the lowest RMS error on the test set.

\begin{figure}
        \centering
        \includegraphics[width=0.8\linewidth]{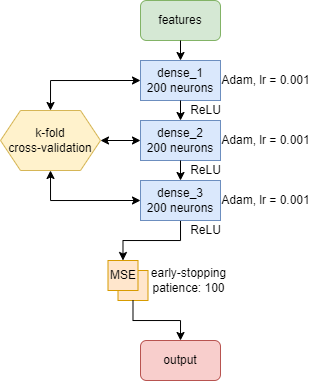}
        \caption[Neural network architecture]{\small Architecture of the neural network used in the NN algorithm. Blue boxes show densely-connected hidden layers, each with 200 neurons and activation functions, and with optimiser and learning rate (lr) indicated. The orange box indicates the loss function (MSE) used to measure the accuracy of the model's training. Arrows indicate the downwards flow of information from one layer to the next. The red box indicates the output layer. Visualisation developed using Bird's Neural Notation Convention \citep{bird2023}.}
        \label{fig:nn_architecture}
    \end{figure}

    As per our previous studies \citep{moss2021_ML}, the NN model used in this study, shown schematically in Figure~\ref{fig:nn_architecture}, was a fully connected NN model using K-fold cross-validation. The NN employed is constructed using the TensorFlow Keras API \citep{abadi2016, tensorflow}, and has a fully-connected, three-layer perceptron architecture designed for regression tasks. Early stopping with a patience of 100 epochs was applied to prevent overfitting. K-fold cross-validation was employed to assess model performance, during which the data in each fold was split into training and test sets. This setup allowed for robust assessment of the model's performance by averaging the evaluation metrics across folds. To manage computational efficiency, each fold was trained over 100 epochs with a batch size of 32. We defined our loss function as mean squared error (MSE) and used mean absolute error (MAE) as the primary evaluation metric. The model was compiled with the Adam optimiser \citep{kingma2017}, which was selected based on prior experiments for its effectiveness in handling sparse and noisy data, which can be problems in astronomical datasets.


\subsection{Comparison with k-Nearest Neighbours}
    Since different ML models can learn different patterns in the same data, as part of the same study we also implemented a kNN algorithm to predict photometric redshifts. The kNN algorithm has been used in previous studies \citep{yuan2013, zhang2013a, zhang2019}. This ML method is computationally less intensive than a NN, as well as being relatively simple and interpretable \citep{zhang2013a}, but it can struggle with complex data and can suffer from catastrophic failures, particularly in certain redshift regimes \citep{han2016}. A NN, on the other hand, can reduce the dispersion and catastrophic outliers, providing more reliable estimates \citep{pasquet-itam2018} at the expense of being computationally more difficult. The kNN's key hyperparameters were optimised and its performance evaluated over 100 iterations. In each iteration, the number of neighbours ($k$) was tested across values from 1 to 40, alongside systematic exploration of distance metrics (Euclidean and Manhattan) and weighting schemes (uniform or distance-based). The combination of these hyperparameters that minimised the RMS error on the test set was identified as optimal, and is listed in Table \ref{tab:knn_hyperparams}. Similar investigations into the impact of distance metrics on kNN-based redshift prediction have been conducted in previous studies, such as \citet{luken2022}, who found that Mahalanobis distance performs best below $z<1$.
    
    Once the optimal hyperparameters were determined for a given iteration, the kNN regressor was retrained on the training set and used to generate predictions for the test set. The residuals $\Delta z = z_\mathrm{spec} - z_\mathrm{phot}$ were calculated, and RMS errors for each $k$ were averaged across all iterations, providing a comprehensive assessment of the algorithm's performance independent of specific data splits.

    \begin{table}
        \centering
        \caption{Final configuration of the kNN model used in this study. These hyperparameters were selected based on grid search performance across 100 iterations, optimising for the lowest RMS error.}
        \begin{tabular}{l l}
            \hline
            \textbf{Hyperparameter} & \textbf{Value} \\
            \hline
            Distance metric & Manhattan \\
            Weighting scheme & distance \\
            Number of neighbours ($k$) & 18 \\
            \hline
        \end{tabular}
        \label{tab:knn_hyperparams}
    \end{table}
\vspace*{\fill}

\section{Results}\label{sec:Results}
    \subsection{DESI fluxes}
        The NN and kNN were trained on the DESI dataset only, using the $g$, $r$, $z$, $W1$ and $W2$ fluxes as training features (see Section~\ref{sec:Preprocessing}). Table~\ref{tab:DESI_performance_metrics} summarises the performance metrics, demonstrating the base level of accuracy we can achieve without invoking wavebands from other surveys.
    
\begin{figure}
            \centering
                \centering
                \includegraphics[width=1.2\linewidth]{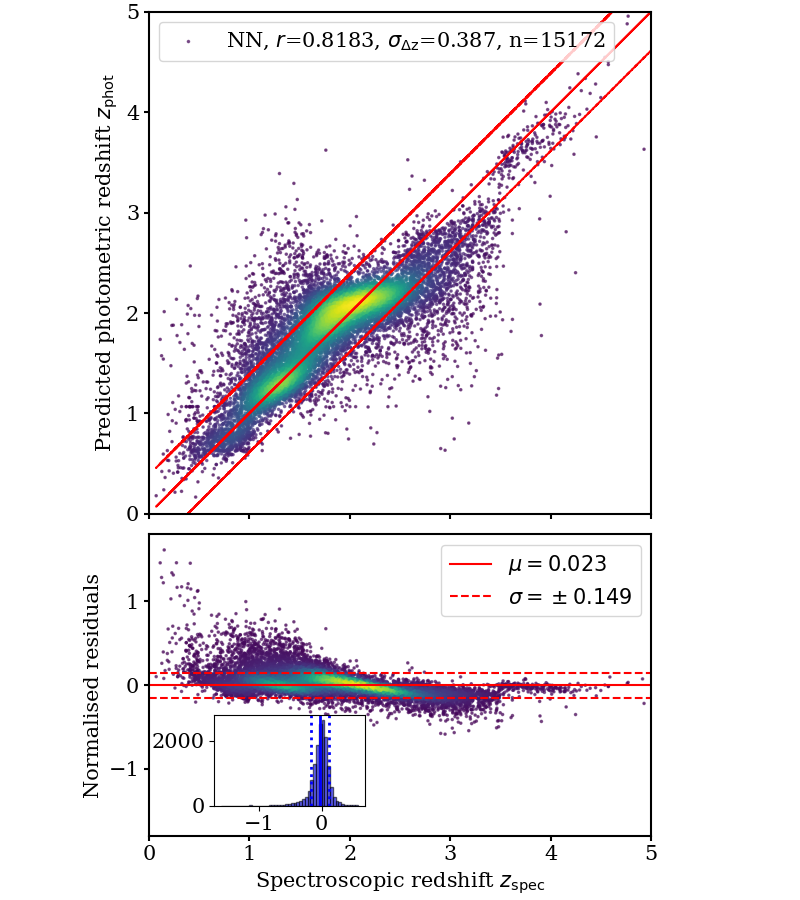}
                \caption{Example plot showing prediction results after training the NN model on the DESI fluxes ($g$, $r$, $z$, $W1$, $W2$). The top panel illustrates the NN predictions, while the bottom panel shows the normalised residuals. In each plot, the solid red line represents the 1:1 relationship, and the dotted red lines indicate the $1\sigma$ deviation from the mean. Inset: distribution of the normalised residuals ($z_\mathrm{phot}-z_\mathrm{spec}$) plotted against redshift, with the red line indicating the line of perfect correlation.}
                \label{fig:desi_nn}
            \hfill
        \end{figure}

        Figure~\ref{fig:desi_nn} shows a representative sample scatterplot for one of the 100 runs of the NN using the DESI fluxes as training features, with the lower panel showing the residuals normalised by $\Delta z/(1+z)$. In this sample, the main cluster of points is near the 1:1 line. Some bimodality is observed in the distribution of points in the scatterplot. This likely reflects complexities in the photometric data, such as the photometric gap between the $z$ and $W1$ wavebands. Importantly, the bimodality does not introduce degeneracies or significantly affect the model's overall performance, as evidenced by the strong correlation coefficient ($r=0.8183$) and low scatter ($\sigma_{\mathrm{\Delta z}} = 0.387$) compared to the kNN model. See Section~\ref{sec:bimodality} for a discussion on how this affects the redshift predictions.

    \subsection{SDSS magnitudes}\label{sec:SDSS}
        \subsubsection{SDSS crossmatching}

\begin{figure}
            \centering
            \includegraphics[width=1\linewidth]{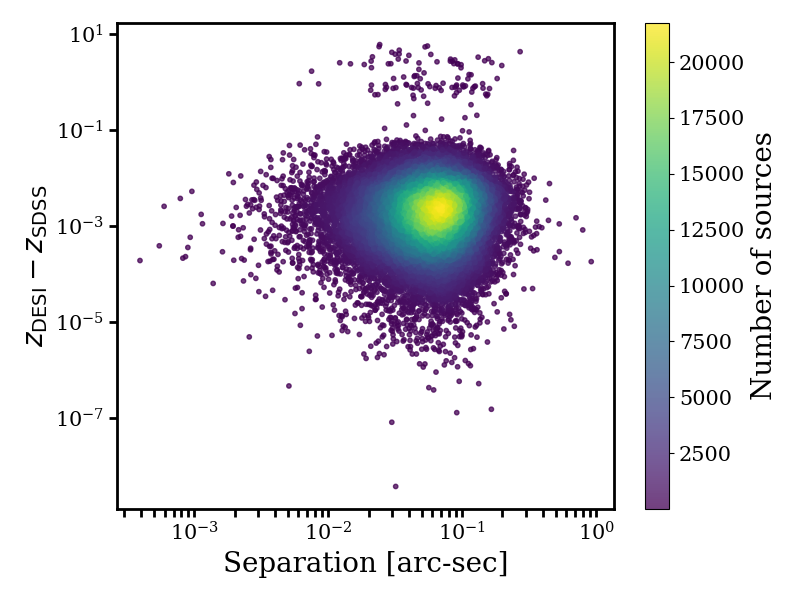}
            \caption{$z_\mathrm{DESI}-z_\mathrm{SDSS}$ versus the angular separation between the DESI and SDSS coordinates.}
            \label{fig:z_vs_separation}
        \end{figure}
\begin{figure*}
           \centering
           \includegraphics[width=0.9\linewidth]{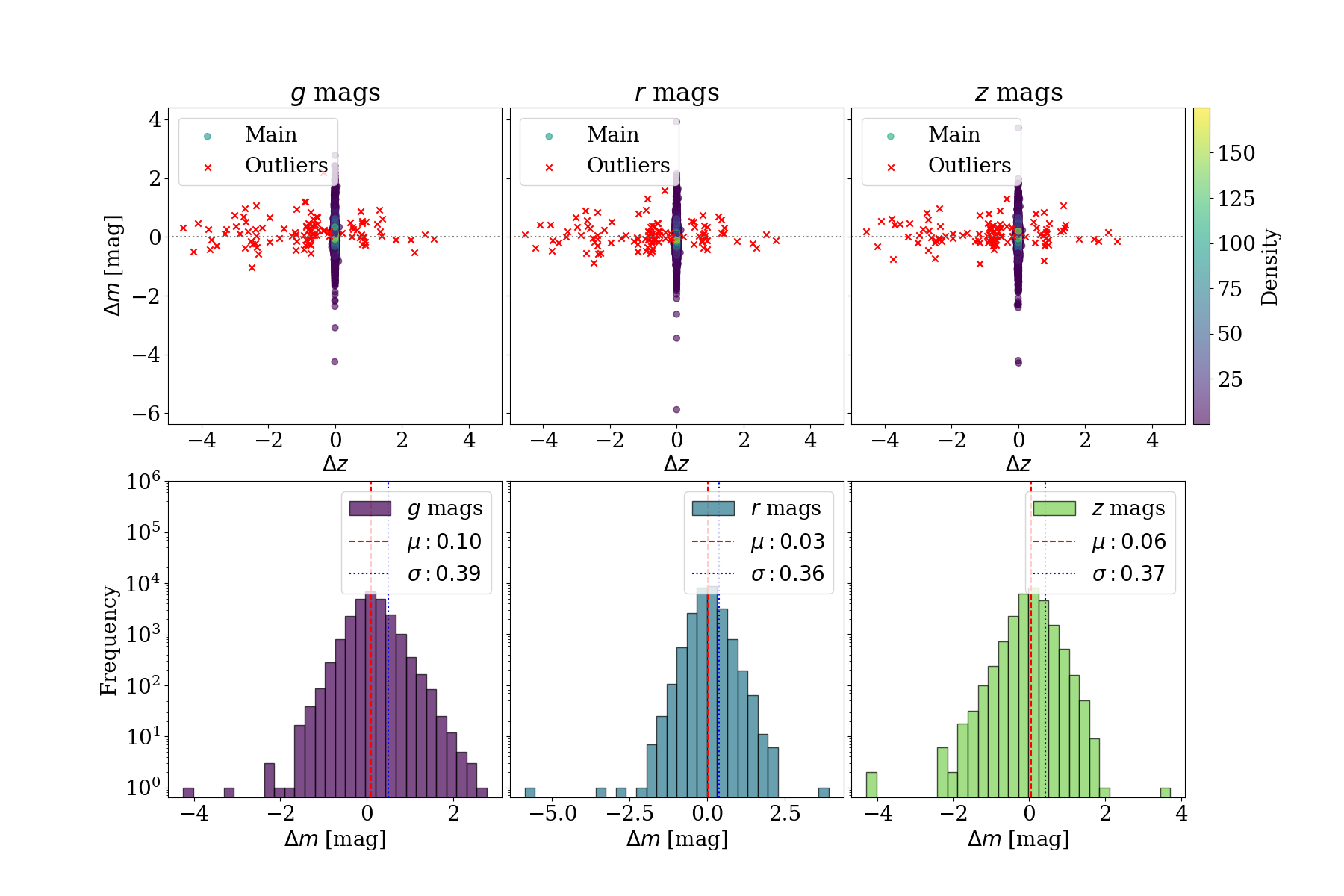}
           \caption{Top: the difference in magnitudes versus difference in redshift between DESI and SDSS measurements, with $m$ being $g$, $r$ or $z$. Red stars show sources for which $z_\text{DESI}-z_\text{SDSS} > 0.14$. Bottom: distribution of the $|\Delta m|$ in the top row.}
           \label{fig:delta_mags_plots}
        \end{figure*}

        The spectroscopic redshifts provided by DESI and SDSS were generally in agreement; however, 107 sources displayed significant discrepancies ($|\Delta z|>0.14$). Such mismatches are most often due to the incorrect association of QSO emission lines \citep{chaussidon2023}. In order to rule out the possibility that these are due to redshift measurements of different sources, Figure~\ref{fig:z_vs_separation} shows the distribution of $|\Delta $z$|$ with the source separation. While there is a grouping of outliers at $|\Delta $z$| \gtrsim0.14$, there is no sign of any correlation. These outliers were excluded from the training sample in order not to contaminate the sample. Lastly, to increase our confidence in the redshifts being measured for the same source, in Figure~\ref{fig:delta_mags_plots} we show $\Delta m$ versus $\Delta z$ for the DxS dataset, with red stars representing the sources for which $|\Delta z| < 0.14 \, (\sim 1 \sigma)$ where $\Delta m$ is the difference between the DESI and SDSS $g$, $r$ and $z$ magnitudes. Although small differences in magnitude for matched DESI and SDSS sources are to be expected, we also considered the impact of excluding sources with large magnitude differences ($|\Delta m|>1$ in $g$, $r$ or $z$). We repeated the analysis both with and without these sources included, and found no significant change in model performance. As a result, only the 107 sources with large redshift discrepancies were excluded from the training set.

        
\begin{figure*}
          \centering
          \begin{subfigure}[t]{0.48\textwidth}
          \raisebox{-0.5mm}{
            \includegraphics[width=\linewidth]{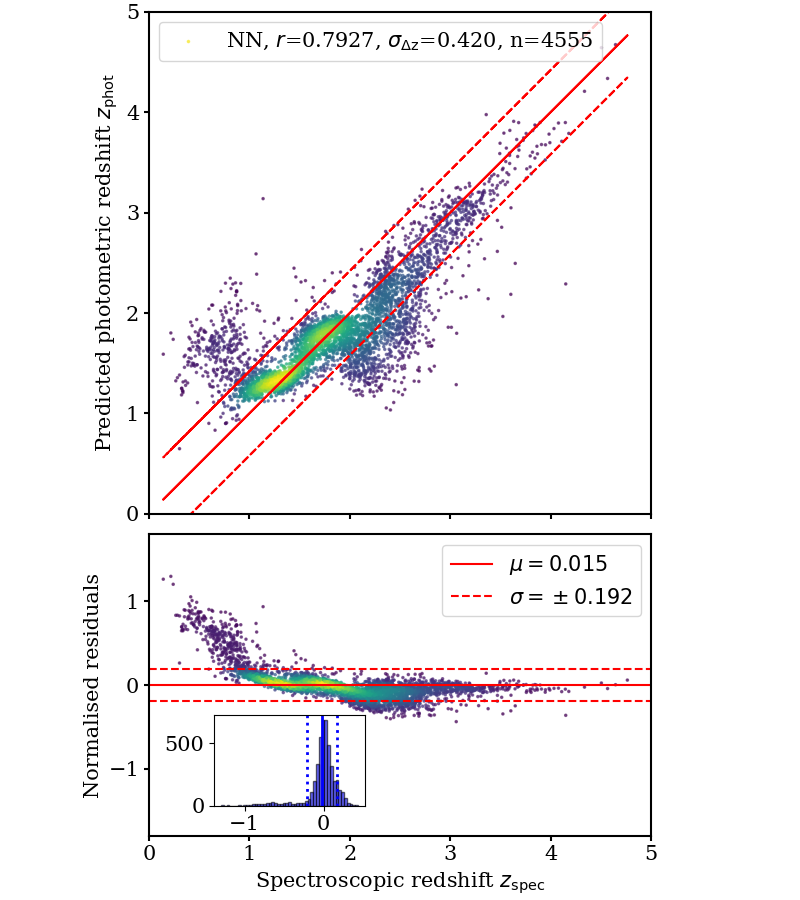}
            }
            \caption{Example plot showing prediction results after training the NN model on the SDSS magnitudes ($u$, $g$, $r$, $i$, $z$). The top panel illustrates the redshift predictions, while the bottom panel shows the normalised residuals. In each plot, the solid red line represents the 1:1 relationship, and the dotted red lines indicate the $1\sigma$ deviation from the mean. Inset: distribution of the normalised residuals ($z_\mathrm{phot}-z_\mathrm{spec}$) plotted against redshift, with the red line indicating the line of perfect correlation.}
            \label{fig:ugriz_nn}
          \end{subfigure}\hfill
          \begin{subfigure}[t]{0.48\textwidth}
            \includegraphics[width=\linewidth]{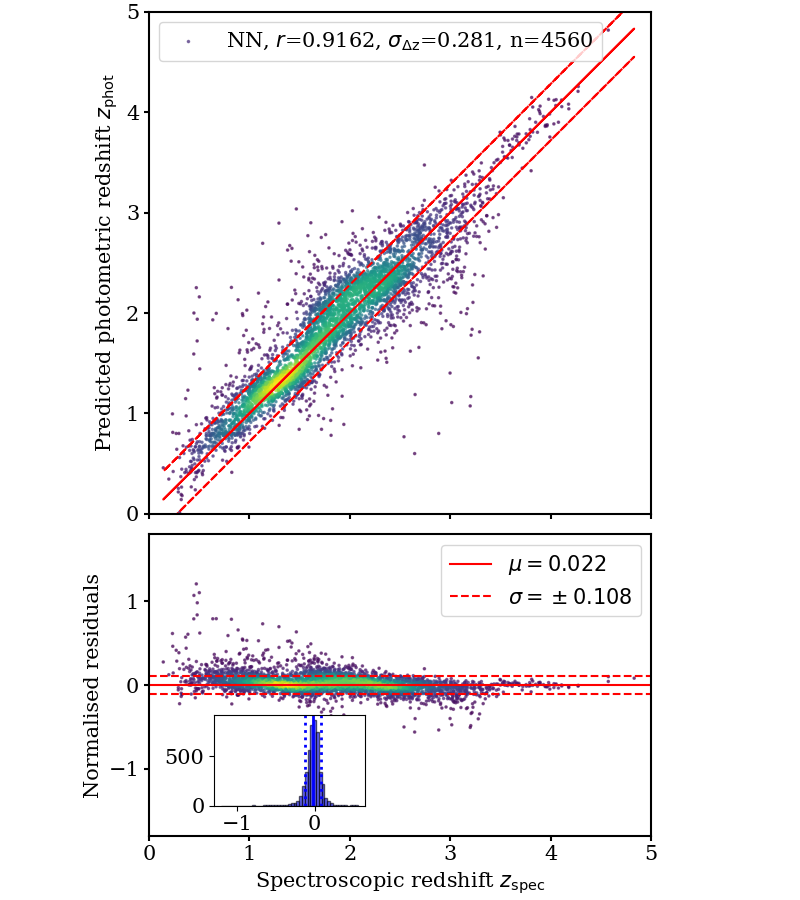}
            \caption{As for Figure~\ref{fig:ugriz_nn} but for the DESI and GALEX fluxes $g$, $r$, $z$, \emph{FUV}, \emph{NUV}.}
            \label{fig:galex_nn}
          \end{subfigure}
          \caption{Comparison of neural network photometric redshift predictions for SDSS-only versus SDSS+GALEX fluxes.}
          \label{fig:nn_comparison}
        \end{figure*}

        The SDSS magnitudes ($u$, $g$, $r$, $i$, $z$) in the DxS dataset were used to train the NN as described in Section~\ref{sec:Preprocessing}, and Table~\ref{tab:SDSS_performance_metrics} gives the averaged metrics across 100 runs of each method. Figure~\ref{fig:ugriz_nn} shows the results of training the NN model on the SDSS magnitudes ($u$, $g$, $r$, $i$, $z$). These predictions are significantly worse than the predictions using either the DESI fluxes ($g$, $r$, $z$, $W1$, $W2$), or DESI complemented with GALEX fluxes ($g$, $r$, $z$, $W1$, $W2$, $NUV$, $FUV$), demonstrating the importance of the $W1$, $W2$, $NUV$ and $FUV$ bands. 

        \begin{table}
            \centering
            \caption{Average performance metrics from 100 runs of the kNN and NN models, trained on the DESI dataset ($g$, $r$, $z$, $W1$, $W2$), where the training and test sets are randomised for each trial. Numbers in parentheses represent uncertainties in the last reported digits. Changes refer to the increase or decrease from kNN DESI to NN DESI. Positive percentage changes indicate an increase (improvement for Corr and EV, but worsening for NMAD, ME, and MAE), and vice versa.}
            \begin{tabular}{lS[table-format=1.4(4)]S[table-format=1.4(4)]S[table-format=2.2]}
            \hline
            \textbf{} & \textbf{kNN DESI} & \textbf{NN DESI} & \textbf{Change (\%)} \\
            \hline
            \textbf{Corr} & 0.7779 \pm 0.0026 & 0.8099 \pm 0.0044 & 3.95 \\
            \textbf{EV} & 0.6046 \pm 0.0041 & 0.6556 \pm 0.0072 & 7.78 \\
            \textbf{NMAD} & 0.2968 \pm 0.0033 & 0.2781 \pm 0.0078 & -6.72 \\
            \textbf{ME} & 3.5971 \pm 0.2949 & 3.0910 \pm 0.4633 & -16.37 \\
            \textbf{MAE} & 0.2002 \pm 0.0022 & 0.1876 \pm 0.0052 & -6.72 \\
            \hline
            \end{tabular}
            \label{tab:DESI_performance_metrics}
        \end{table}
        
        \begin{table}
        \centering
            \caption{As for Table~\ref{tab:DESI_performance_metrics}, but using the SDSS magnitudes ($u$, $g$, $r$, $i$, $z$) only.}
            \begin{tabular}{lS[table-format=1.4(4)]S[table-format=1.4(4)]S[table-format=2.2]}
            \hline
            \textbf{} & \textbf{kNN SDSS} & \textbf{NN SDSS} & \textbf{Change (\%)} \\
            \hline
            \textbf{Corr} & 0.7904 \pm 0.0055 & 0.7770 \pm 0.0373 & -1.72 \\
            \textbf{EV} & 0.6241 \pm 0.0085 & 0.6028 \pm 0.0559 & -3.53 \\
            \textbf{NMAD} & 0.2634 \pm 0.0059 & 0.2968 \pm 0.0431 & 11.25 \\
            \textbf{ME} & 2.2414 \pm 0.2821 & 2.1083 \pm 0.1925 & -6.31 \\
            \textbf{MAE} & 0.1776 \pm 0.0040 & 0.2002 \pm 0.0291 & 11.29 \\
            \hline
            \end{tabular}
            \label{tab:SDSS_performance_metrics}
        \end{table}
        
        \begin{table}
            \caption{As for Table~\ref{tab:DESI_performance_metrics}, but using the DESI fluxes and GALEX $NUV$ and $FUV$.}
            \label{tab:GALEX_performance_metrics}
            \centering
            \begin{tabular}{lS[table-format=1.4(4)]S[table-format=1.4(4)]S[table-format=2.2]}
            \hline
            \textbf{} & \textbf{kNN GALEX} & \textbf{NN GALEX} & \textbf{Change (\%)} \\
            \hline
            \textbf{Corr} & 0.8675 \pm 0.0044 & 0.9187 \pm 0.0036 & 5.57 \\
            \textbf{EV} & 0.7516 \pm 0.0073 & 0.8434 \pm 0.0069 & 10.88 \\
            \textbf{NMAD} & 0.2584 \pm 0.0046 & 0.1971 \pm 0.0047 & -31.1 \\
            \textbf{ME} & 2.4889 \pm 0.1423 & 2.1546 \pm 0.2281 & -15.52 \\
            \textbf{MAE} & 0.1743 \pm 0.0031 & 0.1330 \pm 0.0032 & -31.05 \\
            \hline
            \end{tabular}
        \end{table}

        \begin{table*}
            \centering
            \caption{Performance metrics and relative improvements computed with respect to the NN DESI baseline. Positive percentages indicate an increase (improvement for Corr and EV, but worsening for NMAD, ME, and MAE), and vice versa.}
            \begin{tabular}{l
                S[table-format=1.4(3)] S[table-format=+2.2] S[table-format=+2.2]
                S[table-format=1.4(4)] S[table-format=+2.2] S[table-format=+2.2]}
            \hline
             & \multicolumn{3}{c}{\textbf{NN Model}} & \multicolumn{3}{c}{\textbf{kNN Model}} \\
            \textbf{Metric} & \multicolumn{1}{c}{DESI} & \multicolumn{1}{c}{SDSS cf. DESI (\%)} & \multicolumn{1}{c}{DESI/GALEX cf. DESI (\%)} 
                             & \multicolumn{1}{c}{DESI} & \multicolumn{1}{c}{SDSS cf. DESI (\%)} & \multicolumn{1}{c}{DESI/GALEX cf. DESI (\%)} \\
            \hline
            \textbf{Corr} & 0.8099(44) & -4.23 & +11.84 & 0.7779(26) & +1.58 & +10.33 \\
            \textbf{EV}   & 0.6556(72) & -8.76 & +22.27 & 0.6046(41) & +3.12 & +19.56 \\
            \textbf{NMAD} & 0.2781(78) & +6.30 & -41.1 & 0.2968(33) & -12.68 & -14.86 \\
            \textbf{ME}   & 3.0910(4633) & -46.61 & -43.46 & 3.5971(2949) & -60.48 & -44.53 \\
            \textbf{MAE}  & 0.1876(52) & +6.29 & -41.05 & 0.2002(22) & -12.73 & -14.86 \\
            \hline
            \end{tabular}
            \label{tab:dataset_improvements}
            \end{table*}

    \subsection{GALEX magnitudes}
        As noted previously \citep{Ball2008, niemack2009, Curran2020, nakazono2024}, the inclusion of GALEX narrowband fluxes from DxS alongside DESI fluxes in the second set of features leads to notable improvements in the performance of the machine learning models for redshift predictions (Table~\ref{tab:GALEX_performance_metrics}), the GALEX bands being required to span the $\lambda_{\text{rest}}=1216$ \r{AA} Lyman break at low redshift (see Figure~\ref{fig:variation_of_wavelength_with_redshift}). The kNN and NN were trained on the $g$, $r$, $z$, $W1$, $W2$, $NUV$ and $FUV$ fluxes from DxS. The $NUV$ and $FUV$ measurements come from the SDSS catalogue (see Section~\ref{sec:SDSS}). The inclusion of the $NUV$ and $FUV$ fluxes from the DxS dataset resulted in significant performance improvements for both of the NN and the kNN models. As shown in Table~\ref{tab:GALEX_performance_metrics}, the NN achieves an average correlation across the 100 runs of $r=0.9187\pm0.0036$, a 5.57\% increase over the kNN's correlation of $r=0.8675\pm0.0044$. Substantial improvements are also observed in other metrics: the EV improves by 10.88\%, while the NMAD and MAE decrease by 31.1\% and 15.52\% respectively, indicating higher accuracy and lower scatter in the NN predictions. These results demonstrate the neural network's superior capacity to model the complex relationships in the dataset, using the additional UV fluxes for improved $z_\mathrm{phot}$ predictions. Figure~\ref{fig:galex_nn} shows the results of including the $NUV$ and $FUV$ fluxes from GALEX in the training features for the NN model.

        Table~\ref{tab:dataset_improvements} shows the improvements in metrics across the models and datasets shown in Tables~\ref{tab:DESI_performance_metrics}, \ref{tab:SDSS_performance_metrics} and \ref{tab:GALEX_performance_metrics} to highlight the improvement when moving from one dataset to the next. Training the kNN model on GALEX fluxes as well as DESI fluxes also improves the metrics, but to a much lesser degree than the NN model. The NN trained on the DESI fluxes ($g$, $r$, $z$, $W1$ and $W2$) is used as the baseline for each comparison, and the percentage scores are the changes shown in each metric when the SDSS magnitudes are used as training features, and when the GALEX fluxes are added to the DESI fluxes, respectively. The improvements are calculated as
        \begin{equation}
            \frac{\text{Value}_{\rm new} - \text{Value}_{\rm NN~DESI}}{\text{Value}_{\rm NN~DESI}} \times 100.
        \end{equation}
        Comparing the metrics from the NN model trained on SDSS to those trained on DESI, all metrics are worse with the exception of ME, which is 46.61\% lower when using the SDSS magnitudes as training features. When including the GALEX fluxes along with the DESI fluxes, all metrics improve by $\sim 40\%$, with the exception of correlation coefficient and EV, which show improvements of 11.84\% and 22.27\% respectively.

        To ensure a fair comparison, we tested photometric redshift performance on the same DxS dataset using two sets of features: one with all SDSS bands plus WISE ($u$, $g$, $r$, $i$, $z$, $W1$, $W2$), and one with just the DESI bands ($g$, $r$, $z$, $W1$, $W2$). While the full $ugrizW1W2$ set gave slightly better results, shown in Table~\ref{tab:ugriz_vs_grz}, the differences are small. This suggests that the addition of $u$ and $i$ provides only a modest improvement over $grz+W1W2$. The NIR information from the WISE bands appears to play a larger role in breaking colour-redshift degeneracies than the inclusion of the $u$ and $i$ bands alone.

        \begin{table*}
            \small
            \centering
            \caption{Photometric redshift performance on the same DxS sample ($0.1 < z \leq 4.8$) using two feature sets: $ugriz+W1W2$ and $grz+W1W2$. Metrics are from the kNN model. The addition of $u$ and $i$ yields modest improvements, while WISE bands appear to play a key role in breaking colour–redshift degeneracies.}
            \begin{tabular}{lccccc}
                \hline
                Features & $r$ & EV &NMAD & ME &MAE \\
                \hline
                $ugriz$+$W1W2$ & 0.8439 (0.0038) & 0.7116 (0.0061) & 0.2990 (0.0024) & 2.3504 (0.2105) & 0.2017 (0.0016) \\
                $grz$+$W1W2$   & 0.8329 (0.0030) & 0.6934 (0.0049) & 0.2821 (0.0055) & 2.6612 (0.3839) & 0.1903 (0.0037) \\
                \hline
            \end{tabular}
            \label{tab:ugriz_vs_grz}
        \end{table*}

\section{Discussion}\label{sec:Discussion}
    \subsection{Photometric Differences Between Bimodal Groups}\label{sec:bimodality}
\begin{figure}
            \centering
            \includegraphics[width=1\linewidth]{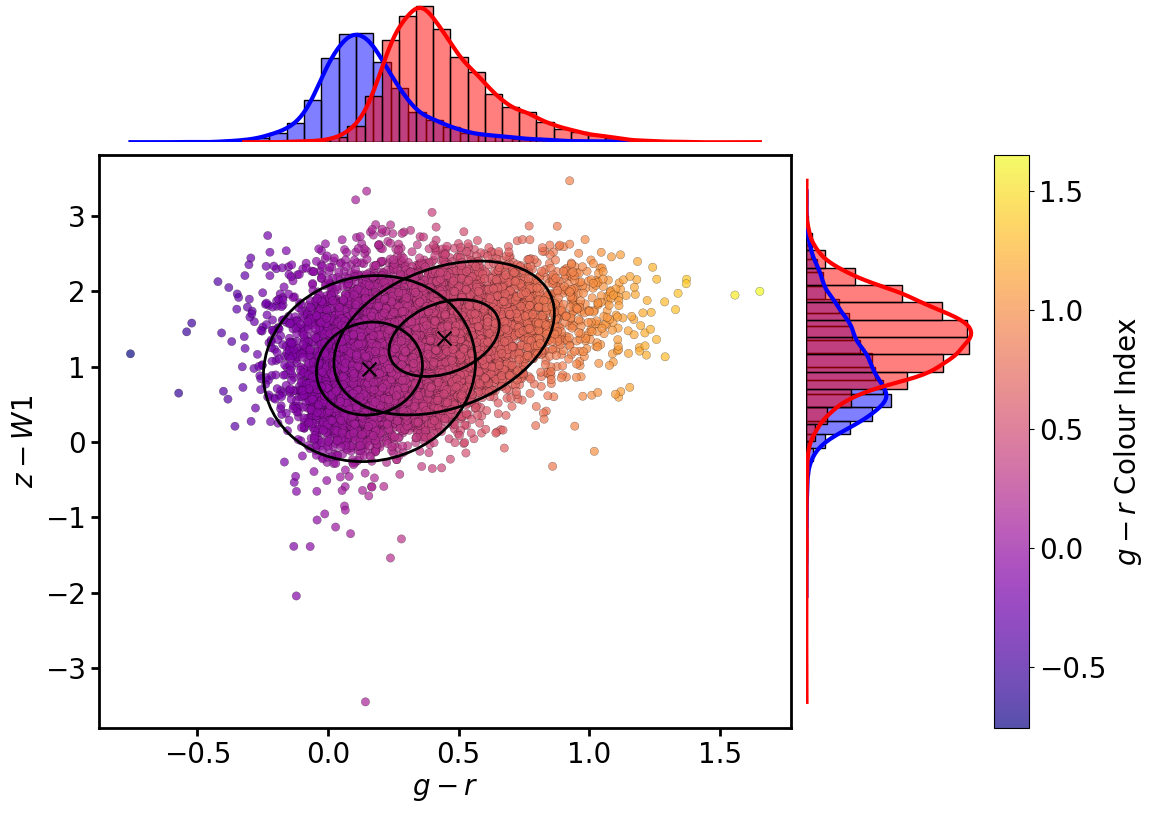}
            \caption{The points from each of the bimodal groups in Figure~\ref{fig:desi_nn} in the $z - W1$ vs. $g - r$ space, incorporating Gaussian ellipses (black ellipses) with centres marked as crosses. The marginal histograms illustrate the distributions of $g - r$ and $z - W1$ within each group, and individual points are coloured by their $g - r$ values, with bluer values on the left and redder values on the right.}
            \label{fig:bimodal_scatterplot}
        \end{figure}

        To investigate the nature of the bimodality observed in the photometric redshift predictions, we examine the differences in photometric colours between the two identified groups. Figure~\ref{fig:bimodal_scatterplot} presents a visualisation of the bimodal groups in colour-space. The bimodal groups were identified by assigning each source a label based on its position in colour space, and the centres of these groups were defined as the mean of their respective colour indices.
    
        To quantify the photometric separation between these two groups, we calculate the Euclidean distances between the mean colour indices (i.e. the centroids) of the bimodal groups in various photometric spaces. Specifically, for each pair of colour indices (e.g., $g - r$ vs. $z - W1$), we calculated the mean values for each group and then measured the Euclidean distance between these two centroids, giving:
        \begin{itemize}
            \item $g - r$ vs. $z - W1$: 0.4964
            \item $g - r$ vs. $r - z$: 0.3355
            \item $z - W1$ vs. $W1 - W2$: 0.4119
            \item $g - r$ vs. $W1 - W2$: 0.2945
        \end{itemize}
    
        These values indicate that the most significant separation between the two bimodal groups occurs in the $g - r$ vs. $z - W1$ space. This suggests that the optical-to-mid-infrared colour combination plays a critical role in distinguishing between the two populations. The relatively large separation in $z - W1$ and $W1 - W2$ further implies that mid-infrared properties contribute to the bimodality, potentially linked to differences in dust obscuration or QSO evolutionary stages.
    
        While the mid-infrared separation ($z - W1$ vs. $W1 - W2$) remains substantial, its slightly lower value compared to the optical-to-mid-infrared colour separation suggests that the primary driver of bimodality is not solely dust reddening. If dust were the dominant factor, we might expect a stronger distinction in the  $g - r$ vs. $W1 - W2$ space (currently the weakest separation at 0.2945).
    
        Furthermore, the large gap between the $z$ and $W1$ photometric bands (see Figure~\ref{fig:variation_of_wavelength_with_redshift}) may be a key factor contributing to the observed bimodality. This gap limits the continuous coverage of spectral features, potentially leading to systematic biases in colour-based classification. The separation in this space suggests that certain populations of QSOs may preferentially occupy distinct regions due to their intrinsic properties or observational selection effects.
    
        The observed photometric separation between the bimodal groups has direct implications for photometric redshift predictions. Since the bimodality is strongest in the optical-to-mid-infrared colour space, it is likely that differences in $z - W1$  play a role in introducing systematic deviations in redshift estimates. If the two groups correspond to distinct physical populations—such as blue, unobscured QSOs versus dust-reddened ones—standard machine learning models may struggle to provide accurate redshifts across the entire QSO sample. This suggests that incorporating explicit bimodal modelling or separate training strategies for these two populations could enhance redshift estimation accuracy.


        The observed bimodality in our $z_\mathrm{phot}$ predictions reflects known degeneracies in broadband photometric data, and is consistent with findings reported by \cite{KuglerGP16} and \cite{disanto2018}. \cite{KuglerGP16} argue that redshift estimation from photometry is fundamentally a multimodal problem, as multiple redshifts can plausibly explain the same set of observed magnitudes due to physical overlaps in SEDs and limited observational constraints. \cite{disanto2018} similarly report strong multi-modal behaviours in photometric redshift predictions across redshift ranges $z \sim 0.5 - 0.9$ and $z \sim 1.5 - 2.5$, attributing this to degeneracies introduced by the use of broadband filters. This phenomenon, though not widely discussed in the context of QSO redshift prediction, may point to important structure in the training data and warrants further study.
        
        
    \subsection{Redshift Dependence}
        \subsubsection{Limitations of the model}
            Figure~\ref{fig:variation_of_wavelength_with_redshift} illustrates how the observed-frame filters ($FUV$, $NUV$, $g$, $r$, $z$, $W1$, $W2$) trace rest-frame wavelengths as a function of redshift. At low redshift ($z\leq 1$), the observed $g$, $r$, and $z$ bands capture rest-frame near-UV to optical wavelengths, which include prominent emission lines such as Mg\,{\sc ii}. At higher redshifts ($z > 2$), these wavelengths are shifted into the $W1$ and $W2$ bands, with UV and optical features moving further into the infrared. This continuous shifting of rest-frame wavelengths across observed bands complicates photometric redshift estimation, particularly at higher redshifts where rest-frame UV features dominate.
            To investigate how each of the ML models operates across different redshift regimes, the DESI dataset was divided into the following redshift bins:
            \begin{itemize}
                \item $0.0 < z \leq 1.0$ (\num{11419} sources)
                \item $1.0 < z \leq 2.0$ (\num{45627} sources)
                \item $2.0 < z \leq 3.0$ (\num{25878} sources)
                \item $3.0 < z \leq 4.0$ (\num{4120} sources)
                \item $4.0 < z \leq 5.0$ (\num{260} sources)
                \item $5.0 < z \leq 6.0$ (\num{13} sources)
            \end{itemize}
            The last redshift range ($z > 5.0$) was disregarded due to the very low number of sources. While the $4.0 < z \leq 5.0$ bin also contains relatively few sources (260), it was retained in the analysis to ensure coverage of the highest redshift regimes included in the dataset.
            
            In the lowest-redshift bin ($0<z\le1$) the addition of the GALEX bands leads to a modest decline in performance: the correlation and EV metrics both dip slightly, while NMAD and MAE increase by a small amount. The maximum error improves only marginally. In the next bin ($1<z\le2$) the two models are effectively indistinguishable once the error bars are accounted for. Beyond $z\approx2$ all metrics worsen for both models, a trend driven by the dwindling number of high-redshift training data and by the fact that rest-frame UV features have shifted into the infrared, where WISE photometry carries larger uncertainties.

            Only a small fraction of DESI QSOs have reliable GALEX detections, and those that do are already among the brighter, better-constrained subset that optical–infrared colours describe well. The GALEX measurements themselves carry comparatively large photometric errors, and therefore, because our current kNN distance metric treats all features equally, the noisy UV magnitudes can blur, rather than sharpen, the $z_\mathrm{phot}$ estimates.
            
\begin{figure*}
              \centering
              \begin{subfigure}[t]{0.48\textwidth}  
              \raisebox{-0.5mm}{
                \centering
                \includegraphics[width=\linewidth]{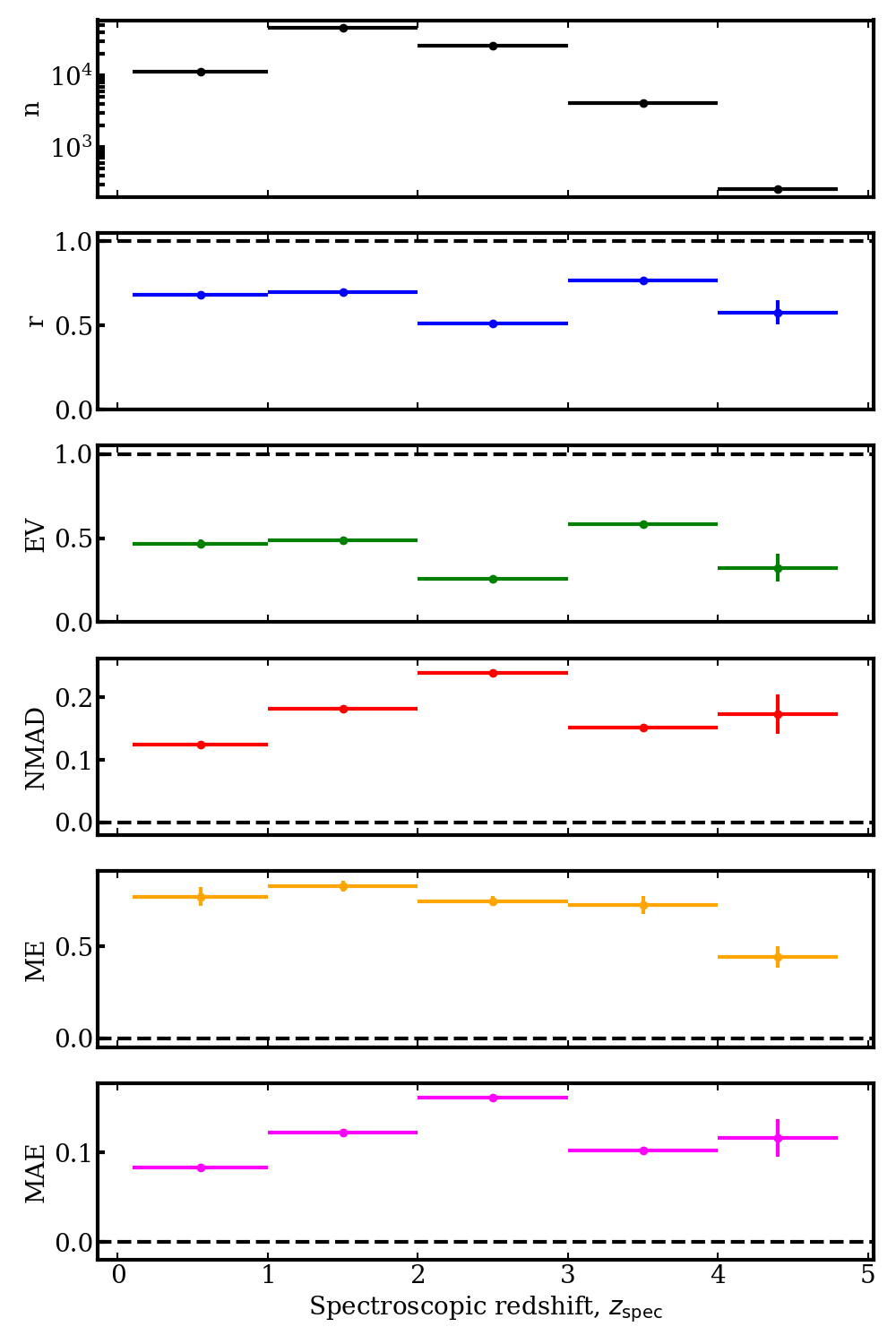}
                }
                \caption{The average of the performance metrics for 100 runs of the kNN model for the DESI ($g$, $r$, $z$, $W1$, $W2$) sample across the five redshift bins. Ideal values for each metric are represented by horizontal dashed lines.}
                \label{fig:DESI_kNN_metrics_plot}
              \end{subfigure}\hfill
              \begin{subfigure}[t]{0.48\textwidth}  
                \centering
                
                \includegraphics[width=\linewidth]{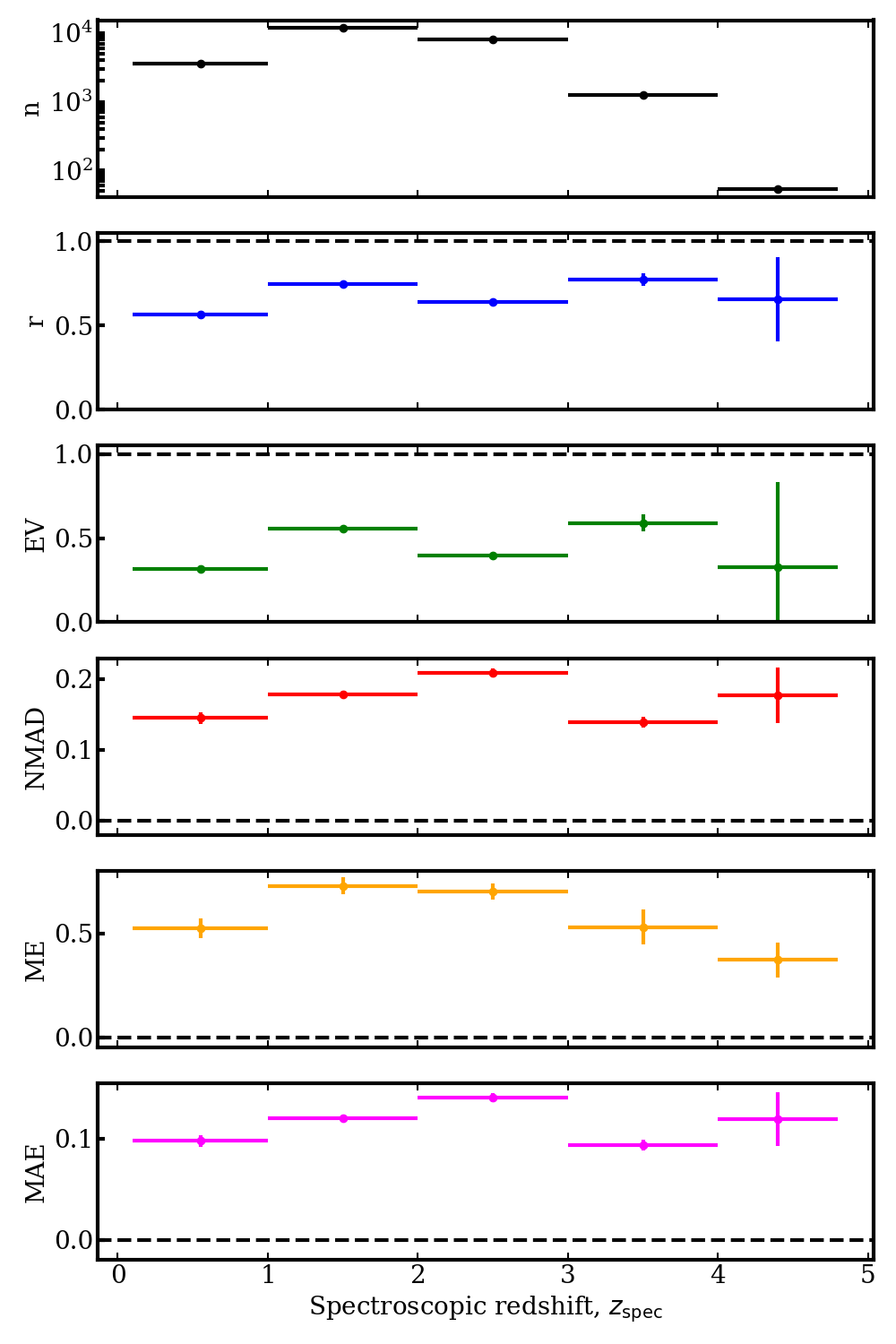}
                \caption{As for Figure~\ref{fig:DESI_kNN_metrics_plot} but for the DxS ($g$, $r$, $z$, $W1$, $W2$, $NUV$, $FUV$) sample.}
                \label{fig:DxS_metrics_plot}
              \end{subfigure}
              \caption{Comparison of kNN performance metrics across the DESI-only and DESI+GALEX (DxS) samples over five redshift bins.}
              \label{fig:kNN_metrics_comparison}
            \end{figure*}

            For the present kNN implementation, Figures~\ref{fig:DESI_kNN_metrics_plot} and \ref{fig:DxS_metrics_plot} show that the inclusion of GALEX $FUV$ and $NUV$ fluxes does not yield a statistically significant improvement in the photometric redshift accuracy, even in the $z\le1$ bin where one might expect the largest benefit. Given the small fraction and relatively poor precision of GALEX detections, this is consistent with the data quality and sampling limitations rather than a failure of the underlying method.

    \subsection{Including GALEX photometry in the training set}

        Previous studies have shown that UV data is especially valuable for blue galaxies and quasars, where traditional optical bands may lack sufficient information to constrain redshifts effectively \citep{niemack2009, zhang2010}. Augmenting the DESI fluxes with GALEX photometry markedly improves our model performance. Both the kNN and NN models benefit from the extended wavelength coverage, although the NN model exhibits particularly enhanced accuracy and robustness across the metrics. While the reduction in ME is modest, improvements in correlation, EV, NMAD and MAE are significant.

        When comparing the photometric redshift estimates from DESI and SDSS, the DESI dataset shows a notably tighter correlation and reduced scatter, especially when GALEX fluxes are added to the DESI bands ($g$, $r$, $z$, $W1$, $W2$), particularly for the NN model. This enhanced performance is largely attributed to the near-infrared coverage provided by $W1$ and $W2$, which, when combined with the UV data from GALEX, helps to break colour–redshift degeneracies more effectively. In contrast, the SDSS dataset ($u$, $g$, $r$, $i$, $z$) lacks this near-infrared component, which may contribute to the overall messier appearance of its correlation plot.        
        
    
        \subsection{Predicting redshift for outliers in DESI/SDSS crossmatched sources}
\begin{figure}
                \centering
                \includegraphics[width=1\linewidth]{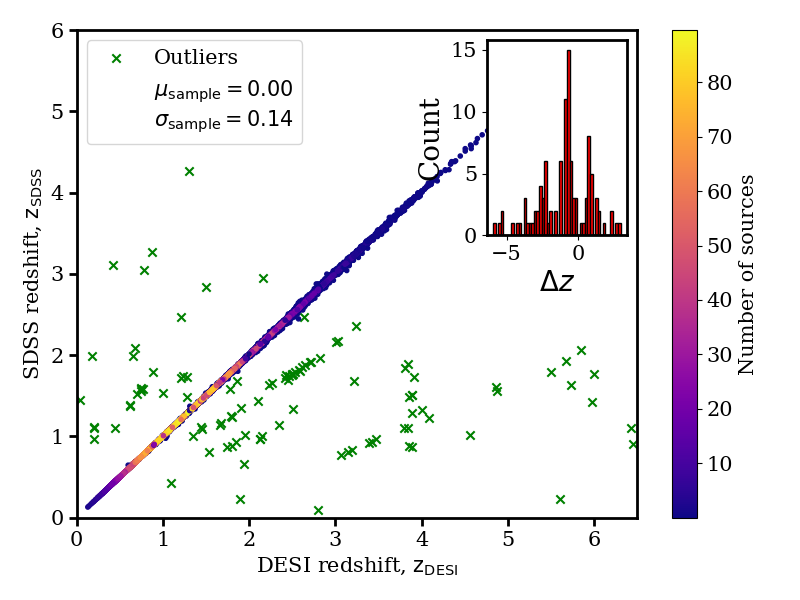}
                \caption[Outliers in $z$ space]{Spectroscopic redshifts from SDSS and DESI for the matched sources in the DxS sample. The background colour-coded scatterplot shows the \num{24509} QSOs for which $z_\text{SDSS} \approx z_\text{DESI}$. The green crosses indicate sources classified as outliers which lie outside $1\sigma \sim 0.14$. The legend shows the mean and standard deviation of the residuals for the full sample. The inset displays the distribution of $\Delta z = z_\mathrm{SDSS} - z_\mathrm{DESI}$ for the outliers only.}
                \label{fig:DxS_outliers}
            \end{figure}

            To identify redshift mismatches, we defined outliers as sources for which $|\Delta z| = |z_\mathrm{DESI} - z_\mathrm{SDSS}| > 0.14$. This threshold corresponds to the $1 \sigma$ width of the residual distribution for the matched sample in Figure~\ref{fig:DxS_outliers}, and was chosen to isolate only the most significant outliers. While some recent studies use $2\sigma$ or $3\sigma$ thresholds (e.g. \citealt{duncan2022,luken2023}), we adopt the more conservative $1\sigma$ definition to ensure a clean separation of the outliers from the dominant population.
            All of these sources come from the DxS crossmatched sample and have complete photometry across all nine bands ($u$, $g$, $r$, $i$, $z$, $W1$, $W2$, $FUV$, and $NUV$), as SDSS provides the optical and UV measurements, and DESI contributes the near-infrared fluxes (see Table~\ref{tab:missing_data_summary}). All nine bands were used as input features in the photometric redshift prediction models for this subset.
    
\begin{figure}
                \centering
                \includegraphics[width=1\linewidth]{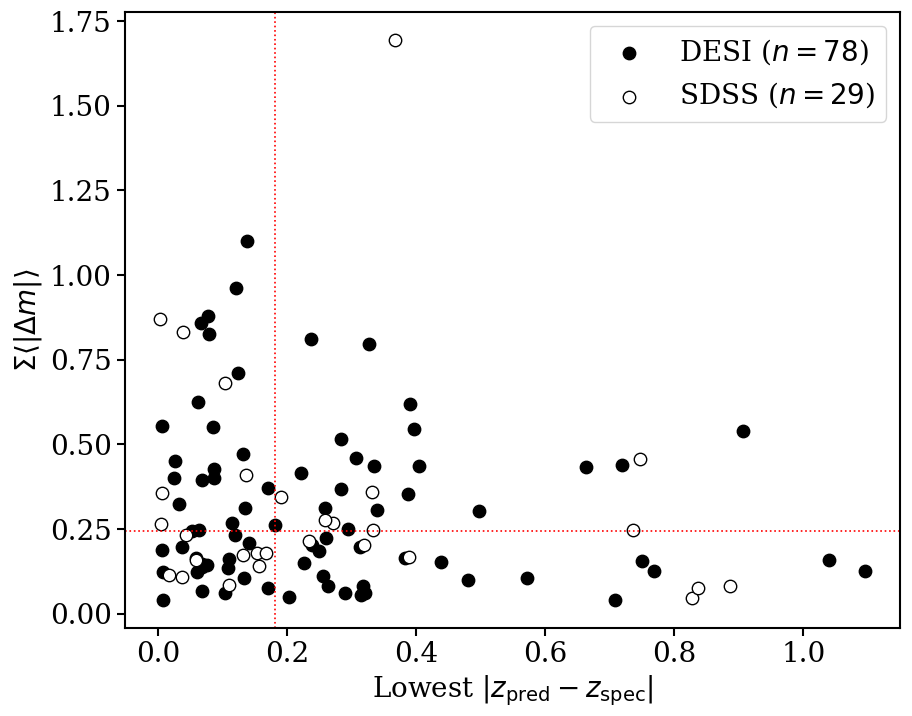}
                \caption{The sum of the difference in the SDSS and DESI $g$, $r$, $z$ magnitudes versus the difference between the predicted and closest spectroscopic redshift for the outliers in Table~\ref{tab:redshift_outliers}. Filled markers show the DESI $z_\mathrm{spec}$ being closest and unfilled for the SDSS. The dotted lines show the median values along each axis, from which we see a concentration at $|\Delta z| \lapprox 0.2$ and a photometric discrepancy of $\langle|\Delta m|\rangle \lapprox 0.6$ in each of the $g$, $r$, $z$ magnitudes.}
                \label{fig:steve's_figure_18}
            \end{figure}

            To address the mismatches in the DESI and SDSS redshifts, we remove the 107 problematic sources and retrain our model on the remaining \num{24509} sources, using the DESI spectroscopic redshift ($z_\mathrm{DESI}$) as the target. Figure~\ref{fig:steve's_figure_18} shows a diagnostic scatterplot comparing the photometric redshift accuracy and photometric consistency for the outlier QSOs, showing that for 78 out of 107 ($\sim73\%$) outliers, our predictions most closely match the DESI spectroscopic redshift. Each point represents an individual source, with the y-axis showing the absolute difference between the SDSS and DESI magnitudes for the source and the x-axis showing the lowest absolute difference between $z_\mathrm{phot}$ and the available spectroscopic redshift from either SDSS or DESI. Objects located in the lower-left quadrant (below both medians) represent cases for which the photometric redshift is relatively accurate and the photometric measurements are consistent between the two surveys. Objects in the upper-right quadrant show larger discrepancies in both redshift and photometry. 
            

\begin{figure}
                    \centering
                    \includegraphics[width=1\linewidth]{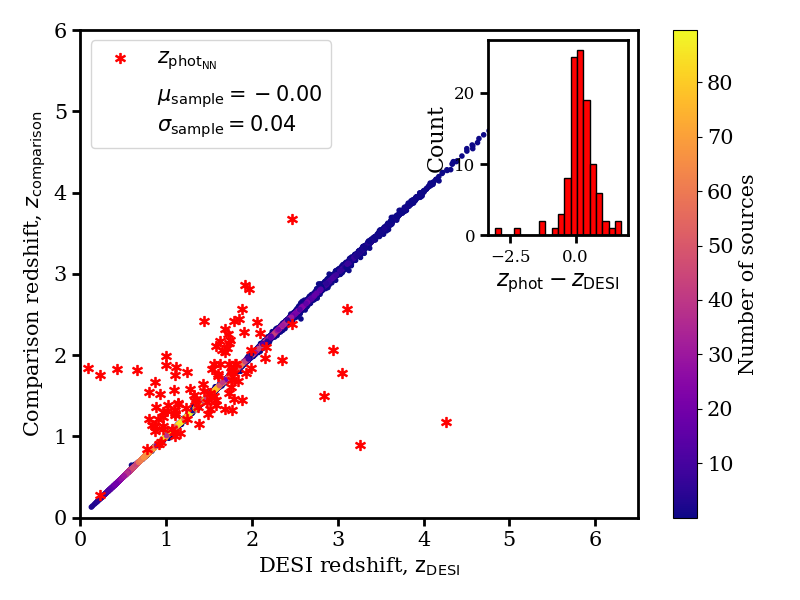}
                    \caption[Outliers in $z$ space]{Photometric redshift predictions from the NN model (red stars) for the outliers identified in Figure~\ref{fig:DxS_outliers}. As in Figure~\ref{fig:DxS_outliers}, the background colour-coded scatterplot shows the \num{24509} QSOs for which $z_\text{SDSS} \approx z_\text{DESI}$. The inset shows the distribution of $\Delta z = z_\mathrm{phot} - z_\mathrm{DESI}$ for these predictions. The predicted redshifts cluster more tightly around the 1:1 line, with improved performance at lower redshifts, especially $1<z<2$ compared to higher redshifts.}
                    \label{fig:outlier_redshift_predictions}
                \end{figure}

            The discrepancies between $z_\mathrm{spec}$ for DESI and those for SDSS were used as an unseen test set for the NN model. In Figure~\ref{fig:outlier_redshift_predictions} we plot the spectroscopic redshifts from SDSS ($z_\text{SDSS}$) against those from DESI ($z_\text{DESI}$). The vertical axis is labelled "$z_{\mathrm{comparison}}$" to reflect that the plotted values may represent either SDSS spectroscopic redshifts or predicted redshifts from our model. For clarity, each data source is explicitly indicated in the figure legends. The red stars show the redshift predictions from our model plotted against the spectroscopic redshift from DESI. Figure~\ref{fig:outlier_redshift_predictions} also shows that the NN model tightens up the $z_\mathrm{phot}$ for the outliers, which is confirmed in the inset, which shows that the standard deviation of the residuals is now $\sigma_\mathrm{resids}=0.04$ (improved from $0.14$). Many of the predicted redshifts are now on the line of $z_\text{SDSS} \approx z_\text{DESI}$.
    
    \subsection{Comparison with previous machine learning results}
        Photometric redshift estimation has been approached using a variety of different techniques, each with distinct strengths and limitations. Broadly speaking, these methods can be categorised into template-fitting, machine learning, and deep learning approaches.

        Template-fitting approaches, such as those using \textsc{eazy} \citep{brammer2008a}, have traditionally been employed for high-redshift sources, where training data for ML models is sparse. \cite{li2022} demonstrate that combining template fitting with ML, using \textsc{catboost} for low redshift galaxies and \textsc{eazy} for extrapolation to high redshifts, improves accuracy, particularly for $z < 2$. However, this approach still struggles with out-of-distribution predictions, reinforcing the need for robust high-redshift solutions.

        Machine learning models have shown promise in improving redshift estimation by training algorithms on large datasets, using feature selection techniques. \cite{saxena2024} introduce CircleZ, a NN model optimised for AGNs, incorporating photometric and morphological features to achieve high precision in $z_\mathrm{phot}$ estimation. Similarly, \cite{hong2022} propose a multimodal ML approach, integrating photometric and spectroscopic features to enhance redshift predictions for QSOs, significantly reducing the RMSE. However, these methods remain sensitive to the quality and representativeness of their training datasets, often struggling with selection biases and incomplete sky coverage.

        Deep learning techniques, such as those implemented by QuasarNET \citep{busca2018}, provide an alternative approach by directly learning spectral features from large spectroscopic datasets. QuasarNET demonstrates near expert-level performance in QSO classification and redshift estimation, using convolutional neural networks to detect emission lines and refine redshift estimates. While this method excels at identifying broad absorption line (BAL) QSOs and reducing catastrophic errors, its reliance on identifiable spectral features limits its applicability to high-redshift QSOs, where fewer lines are available.

        Compared to other recent methods, our NN approach performs competitively while offering greater flexibility across a wider redshift range. While QuasarNet is highly effective for spectroscopic redshift estimation via spectral line identification, it is not directly applicable to purely photometric datasets. In contrast, our method relies solely on photometry, allowing it to be deployed across large sky areas with limited or no spectroscopic follow-up. The use of ML also enables the model to adapt to the diverse spectral energy distributions of QSOs without requiring template tuning or manual line matching.

        Our method builds on these developments by tailoring photometric redshift estimation to the characteristics of QSOs across a wide redshift range. Compared to \cite{saxena2024}, who use optical and infrared (IR) photometry of X-ray selected AGNs from the Legacy Imaging Survey in $g$, $r$, $i$, $z$ and $W1 - W4$, our dataset spans from mid-infrared wavelengths from the WISE to UV from GALEX, offering broader wavelength coverage. While \cite{li2022} successfully apply template fitting to high-redshift galaxies, we evaluate multiple ML models to better address the complexity of QSO spectral energy distributions. Our neural network, a configurable regression model, provides a flexible approach to estimating $z_\mathrm{phot}$ as a continuous variable. While it lacks specialised architectural optimisations for spectral line identification, its adaptability makes it well-suited for photometric redshift estimation. By refining feature selection and expanding wavelength coverage, our study enhances QSO redshift predictions and informs future wide-field survey analyses.

\section{Conclusions}\label{sec:Conclusions}
    Given that a simple, accurate, and reliable photometric estimate of redshift for samples of QSOs will be invaluable for upcoming large radio surveys, we have developed and evaluated a neural network capable of predicting the redshift of QSOs in the Dark Energy Spectroscopic Instrument Early Data Release.

    Our neural network model achieves a correlation coefficient of $r=0.81$ with spectroscopic redshifts, with $NMAD = 0.28$. The inclusion of UV photometry from GALEX improves the redshift predictions to a correlation of $r=0.92$ (a 13\% increase) and NMAD of 0.197 (a 29\% reduction), while also reducing scatter and catastrophic outliers. This improvement is particularly significant for high-redshift QSOs, where rest-frame UV features shift into the optical bands, making UV photometry a valuable addition to redshift estimation models.
    
    A notable feature of our results is the bimodal distribution observed in the photometric redshift predictions, which has not been explicitly discussed in the literature. Our analysis suggests that this is linked to differences in the $g - r$ vs. $z - W1$ colour space, likely arising from systematic biases within our data, or from the large photometric gap between DESI’s $z$ and $W1$ bands. This finding suggests that further refinement of redshift prediction models, potentially incorporating tailored treatments for bimodal populations, could improve accuracy.
    
    We also assess the impact of missing photometric bands compared to SDSS, particularly the absence of the $u$ and $i$ bands. Our results show that the inclusion of the GALEX UV fluxes provides additional constraining power, resulting in comparable or superior performance to models that include SDSS $u$ and $i$ bands. This suggests that deep UV coverage may be preferable to broadband optical coverage for QSO redshift predictions.
    
    Additionally, we investigate mismatches between DESI and SDSS spectroscopic redshifts, finding that DESI redshifts are generally more reliable in cases of significant discrepancies. These mismatches are likely due to incorrect emission line associations in SDSS spectra, reinforcing the accuracy of DESI's spectroscopic redshift measurements.
    
    Unlike many previous studies that focus on galaxies with well-defined spectral features or which cover only narrow redshift ranges, our approach uses a neural network trained on the crossmatched DxS dataset to estimate photometric redshifts for QSOs spanning the full redshift range observed in DESI. As large-scale surveys such as the SKA come online, reliable photometric redshifts will be essential for identifying and analysing large QSO samples without spectroscopic follow-up. By incorporating deep UV photometry, we demonstrate that a machine learning model can effectively capture the spectral diversity of QSOs, improving redshift predictions across a broad range of source types. Future refinements that explicitly model different QSO spectral classes may offer further gains in precision.



\appendix
\section{k-Nearest Neighbours}\label{sec:kNN}
\begin{figure}
    	\centering
    	\includegraphics[width=1\linewidth]{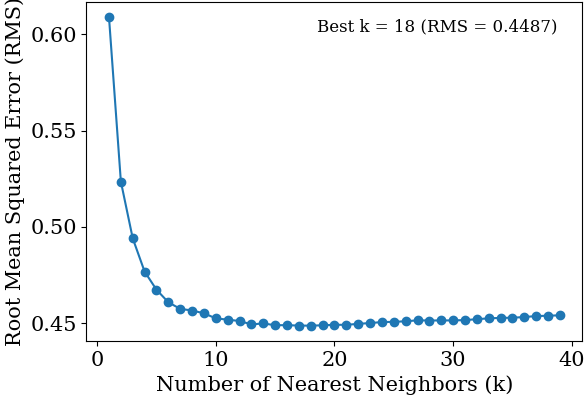}
    	\caption[Best k value]{Root Mean Squared Error (RMS) vs. Number of Nearest Neighbours (k) for the kNN, showing the average RMS error for each value of $k$. The error decreases sharply for small values of k and stabilises around $k=18$, indicating an optimal choice for this parameter.}
    	\label{fig:best_k}
    \end{figure}

    A kNN method was also employed to estimate the redshift of QSOs from photometric data in the DESI dataset as a comparison to the NN. The kNN algorithm iteratively searches for an optimal number of nearest neighbours, measuring the model’s performance using the root mean squared (RMS) error between predicted and actual redshifts. A series of 100 iterations was conducted to obtain a representative optimal number of nearest neighbours, ensuring robustness in the selection process. The metrics in Tables~\ref{tab:SDSS_performance_metrics} and \ref{tab:GALEX_performance_metrics} are averaged across the 100 runs, and their standard deviations calculated as a measure of the uncertainty in each metric.
    Figure~\ref{fig:best_k} illustrates how the MSE changes in a representative run of the kNN algorithm as the number of nearest neighbours in the model varies. The error decreases rapidly for low values of $k$, stabilizes around $k=18$, and then slowly rises again. This pattern reflects the typical balance between over- and under-fitting.


    Following the determination of the optimal $k$ value, the model undergoes further refinement by testing different distance metrics (Euclidean vs Manhattan) and weighting schemes (uniform vs distance). The model is then finalised by selecting the best-performing distance metric and weighting strategy based on the mean squared error (MSE).

    To assess the importance of each of the features (fluxes) in producing redshift predictions, the \texttt{SelectKBest} method, using f-regression, is applied to quantify the contribution of individual photometric features. Additionally, permutation importance is employed, offering a non-parametric measure of feature significance.

\begin{figure}
            \centering
                \centering
                \includegraphics[width=\linewidth,keepaspectratio]{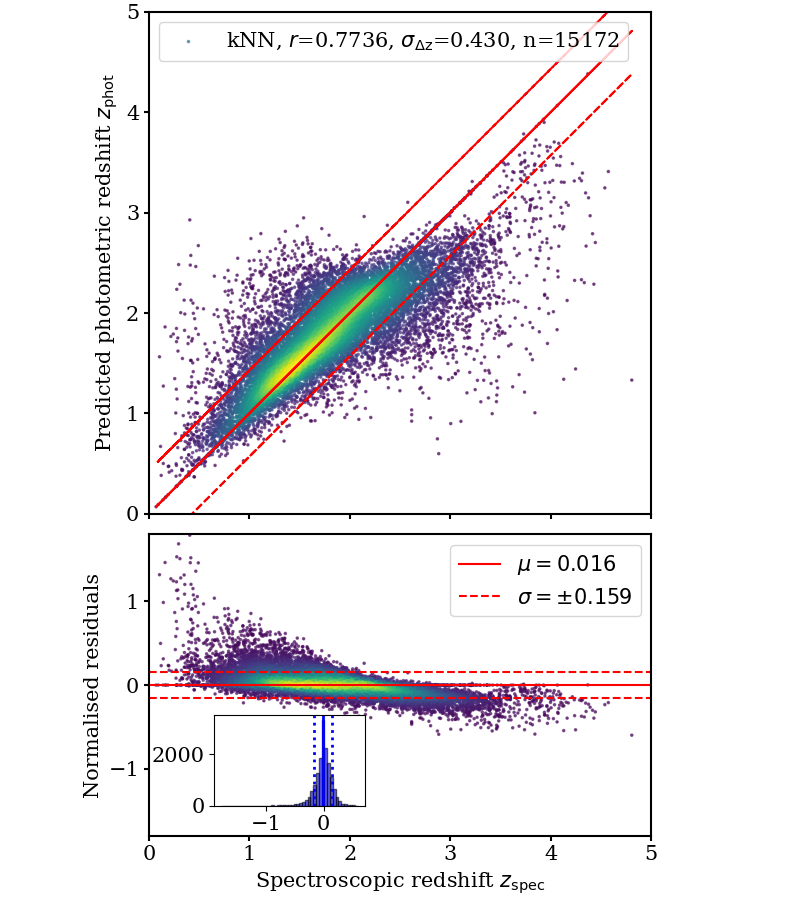}
                \caption{Example plot showing prediction results after training the kNN model on the DESI fluxes ($g$, $r$, $z$, $W1$, $W2$); cf. Figure~\ref{fig:desi_nn}. The top panel illustrates the kNN predictions, while the bottom panel shows the normalised residuals. In each plot, the solid red line represents the 1:1 relationship, and the dotted red lines indicate the $1\sigma$ deviation from the mean. Inset: distribution of the normalised residuals ($z_\mathrm{phot}-z_\mathrm{spec}$), with the mean and standard deviation indicated.}
                \label{fig:desi_knn}
        \end{figure}

\begin{figure}
            \centering
                \centering
                \includegraphics[width=\linewidth,keepaspectratio]{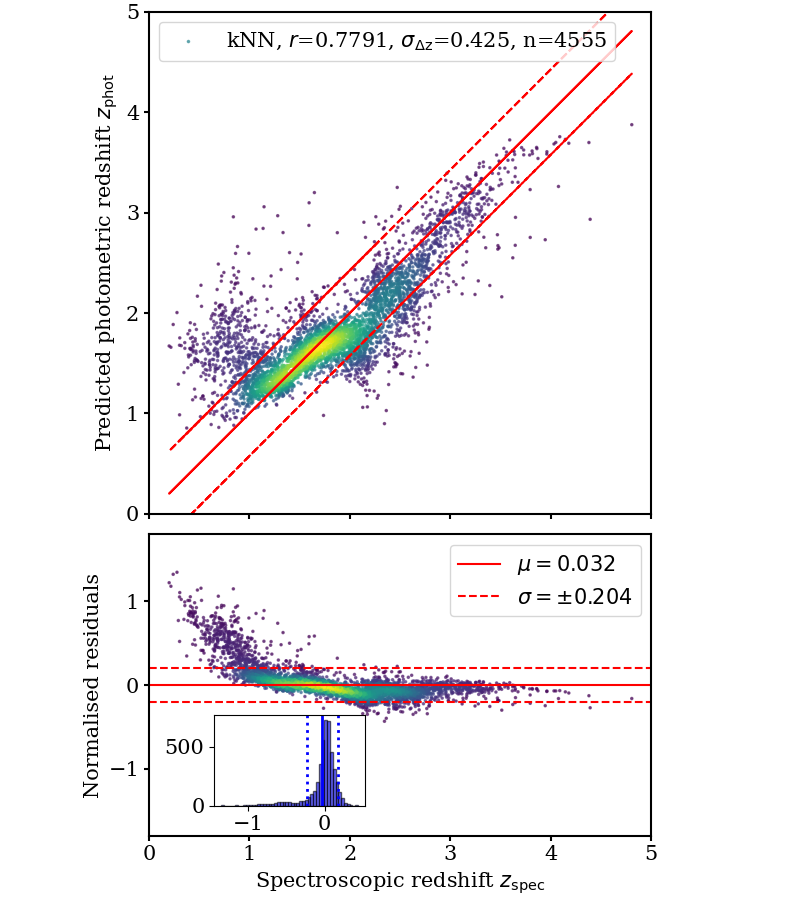}
                \caption{As for Figure~\ref{fig:desi_knn}, but using the SDSS ugriz magnitudes as training features. Cf. Figure~\ref{fig:ugriz_nn}.}
                \label{fig:ugriz_knn}
        \end{figure}

\begin{figure}
            \centering
                \centering
                \includegraphics[width=\linewidth,keepaspectratio]{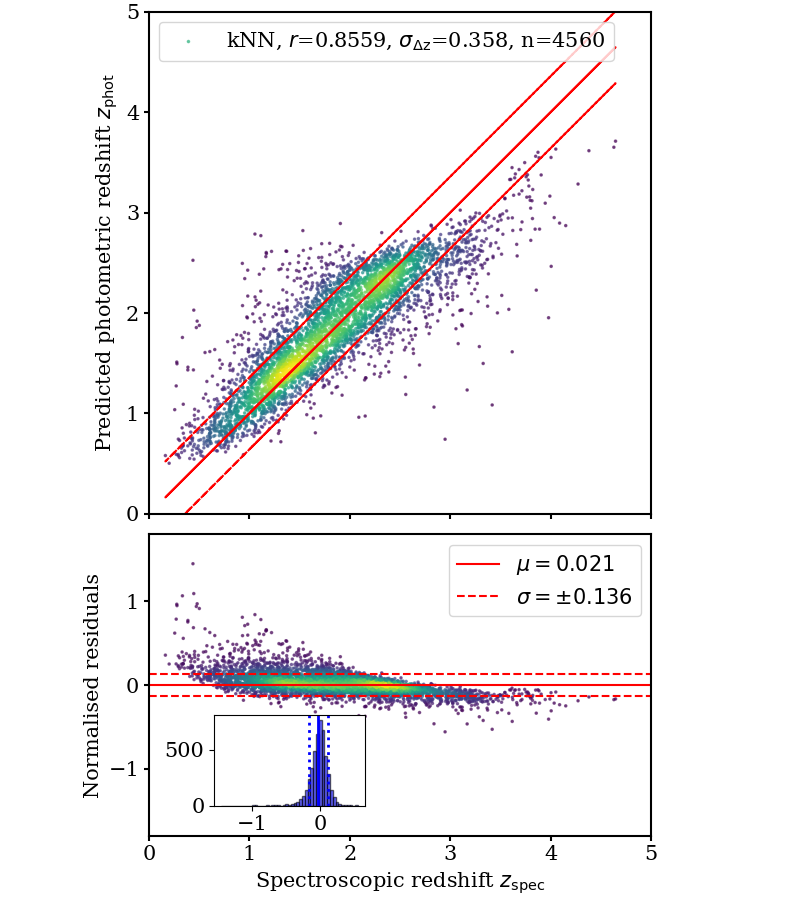}
                \caption{As for Figure~\ref{fig:desi_knn}, but using the DESI and GALEX fluxes together as training features; cf. Figure~\ref{fig:galex_nn}.}
                \label{fig:galex_knn}
        \end{figure}

    Just as Figure~\ref{fig:ugriz_nn} illustrates the correlation between $z_\mathrm{phot}$ and $z_\mathrm{spec}$ for the neural network trained on SDSS magnitudes, Figures~\ref{fig:desi_knn}, \ref{fig:ugriz_knn}, and \ref{fig:galex_knn} present the corresponding results for the kNN model trained on the same features across the same datasets.
        
    The inclusion of GALEX $FUV$ and $NUV$ fluxes significantly improves the accuracy of $z_\mathrm{phot}$ predictions, as depicted in Figure~\ref{fig:galex_knn}, particularly for specific subsets of QSOs.

    \subsection{Feature importance}\label{sec:feature_importance}
    \subsubsection{Permutation importance method}
        Regarding the interpretability of a machine learning model, it is instructive to determine which wavebands in an astronomical survey are most informative for predicting photometric redshifts. For each model, feature relevance was assessed using permutation importance, in which the values of each feature are randomly shuffled to observe the effect on model performance. Features whose randomisation leads to a substantial drop in accuracy are deemed more important. This approach highlights which photometric bands contribute most to determining $z_\mathrm{phot}$, although the results can vary depending on the choice of $k$ and the properties of the training data.

\begin{figure*}
          \centering
          \begin{subfigure}[t]{0.48\textwidth}
            \includegraphics[width=\linewidth]{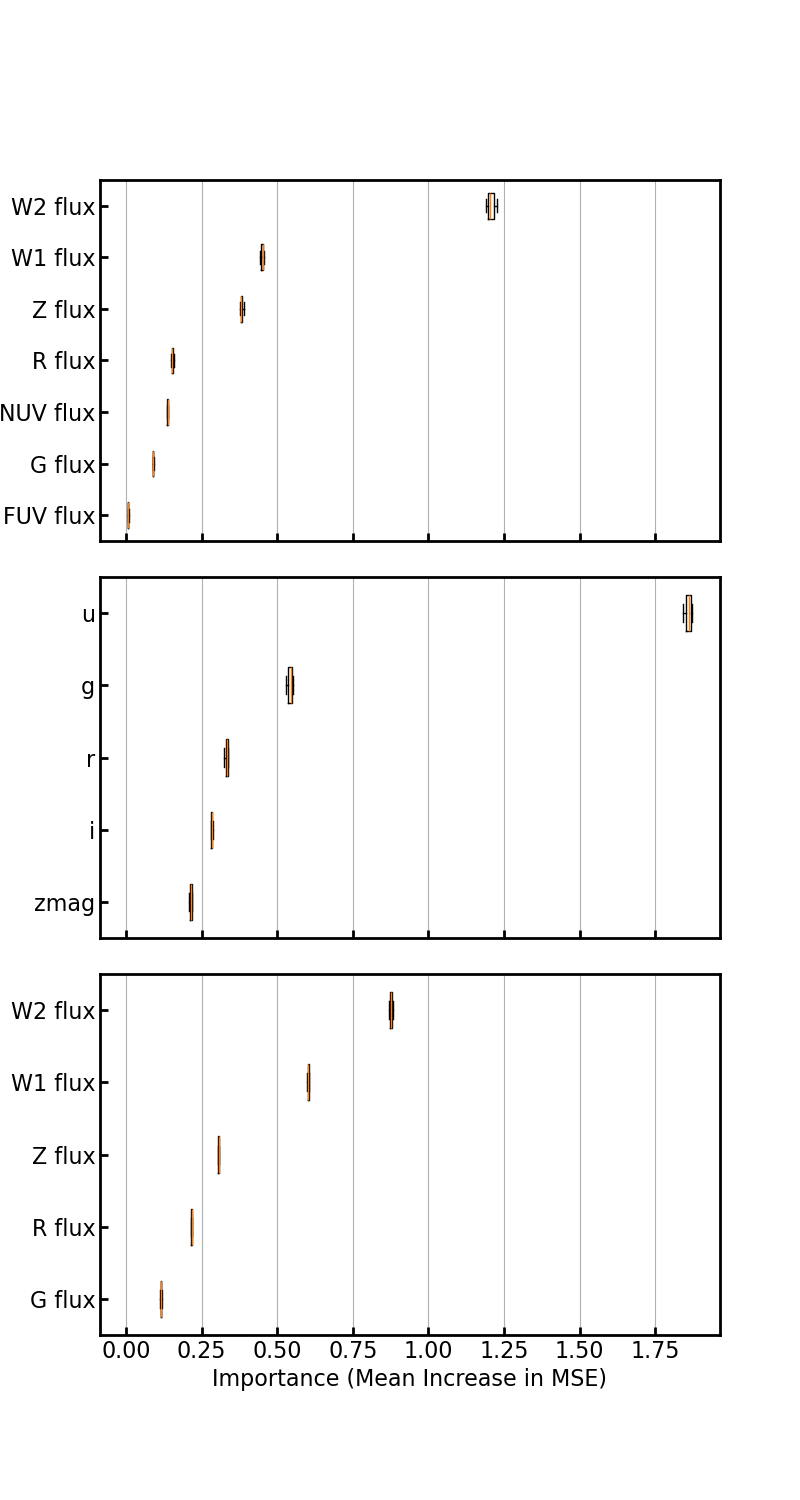}
            \caption{Feature importances for the kNN model on the DESI/GALEX sample.}
            \label{fig:knn_importances}
          \end{subfigure}\hfill
          \begin{subfigure}[t]{0.48\textwidth}
            \includegraphics[width=\linewidth]{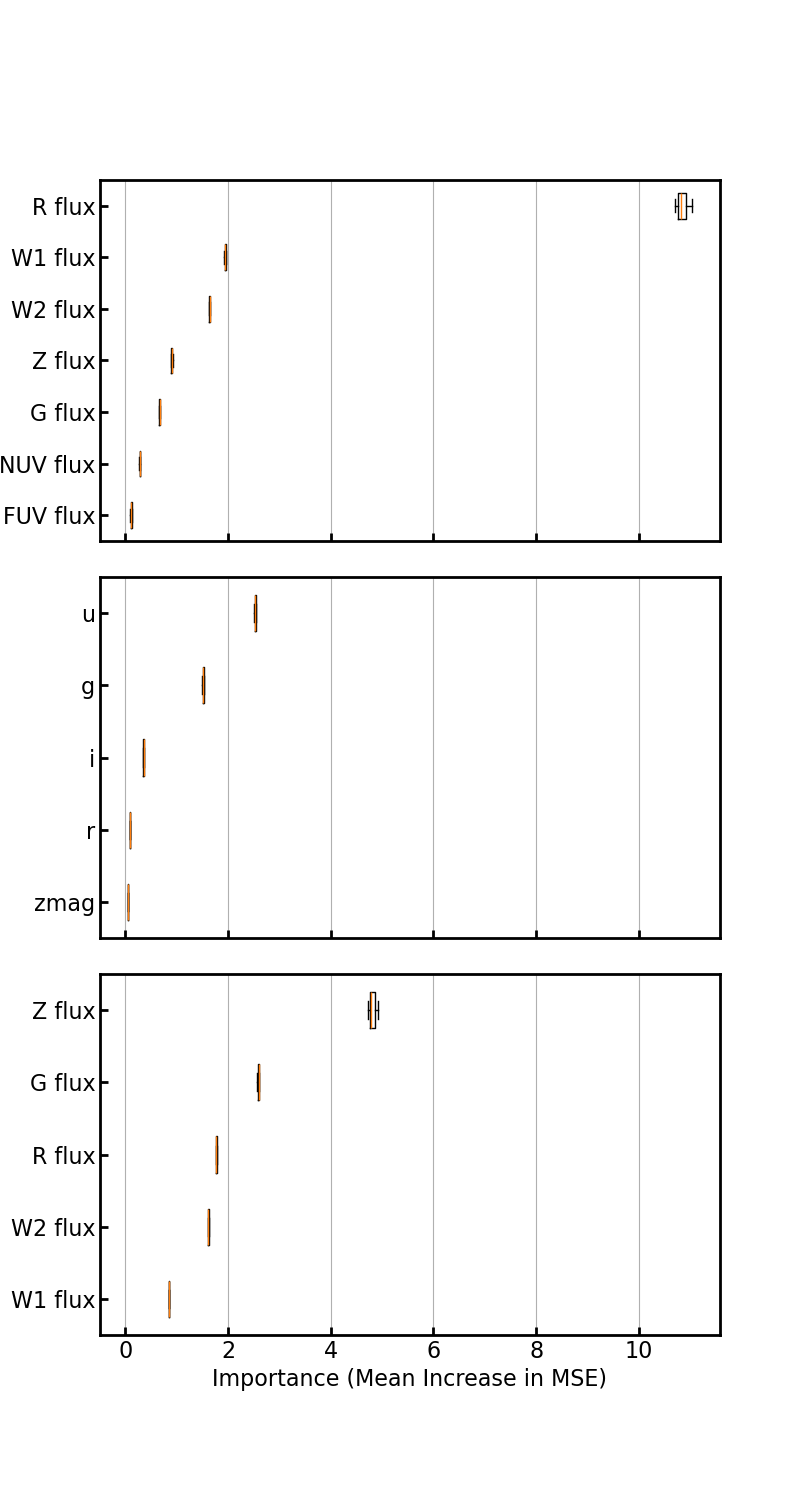}
            \caption{As for Figure~\ref{fig:knn_importances} but for the neural network model.}
            \label{fig:nn_importances}
          \end{subfigure}
          \caption{Feature importances from (a) kNN and (b) neural network models. Each panel shows the mean increase in MSE when omitting each flux in the DESI/GALEX sample.}
          \label{fig:importances_comparison}
        \end{figure*}

        The permutation importances from a representative run of the kNN model, based on the DxS dataset, are shown in Figure~\ref{fig:knn_importances}. The top panel shows the DESI/GALEX fluxes, and the middle and bottom panels show the importances for the SDSS magnitudes and the full DESI sample, respectively. In both cases, the infrared WISE bands were ranked among the highest, indicating that they are more predictive than the optical bands alone, although all features appear to contribute meaningfully.
        
        Feature importances were also evaluated for the neural network model using permutation importance, averaged over 100 runs. The permutation importances for one run of the NN for the same datasets as used above are shown in Figure~\ref{fig:nn_importances}. In the DESI/GALEX dataset (top panel), both the NN and kNN models ranked the $r$, $z$, $W1$, and $W2$ fluxes highest, with $FUV$ and $NUV$ appearing near the bottom. In the SDSS component of DxS, the $u$ and $g$ bands were consistently assigned the greatest importance across both models. These filters likely provide key information due to their sensitivity to strong quasar spectral features, such as the Lyman break at moderate to high redshift.
        
        \subsubsection{Drop-column method}
            The measured importance of $FUV$ and $NUV$ fluxes can depend on the method used. Permutation importance may underestimate the value of features that are highly correlated with others, as the model can compensate using redundant inputs. Consequently, even if $FUV$ and $NUV$ improve overall model accuracy, their permutation scores may not fully reflect their utility.
            
            To better assess the true impact of individual features, we applied the drop-column method, which involves retraining the model after removing each feature entirely. The results for the kNN model are summarised in Table~\ref{tab:drop_column_results_knn}. Removing $FUV$ led to an increase in both $\sigma_\mathrm{NMAD}$ and MAE, indicating a measurable degradation in performance. Similarly, dropping $W1$ or $W2$ degraded model accuracy, confirming their predictive value. By contrast, removing $NUV$ had little effect, suggesting greater redundancy between $NUV$ and the remaining features. These findings highlight the limitations of permutation-based scores in the presence of correlated features.
            
            We also applied the drop-column method to the NN model using the DESI and DxS datasets, as shown in Table~\ref{tab:drop_column_results_nn}. Removing $W1$ and $W2$ fluxes increased both $\sigma_\mathrm{NMAD}$ and MAE, confirming their roles as key predictors. A similar degradation was observed when $NUV$ and $FUV$ fluxes were excluded, despite these features appearing near the bottom of the feature importance rankings. This discrepancy highlights a known limitation of permutation importance: when a feature is correlated with others, or when its contribution is localised to specific subpopulations, randomising it may not strongly impact performance, as the model can partially compensate using other inputs. In contrast, dropping the feature entirely removes its unique contribution, offering a more reliable test of importance in such cases. These results emphasise their significance for accurate photometric redshift predictions in the neural network model.

            Taken together, the results from both permutation and drop-column analyses confirm the critical role of near-infrared and some optical bands, particularly $W1$, $W2$, $NUV$, and $FUV$ fluxes, in accurate $z_\mathrm{phot}$ estimation. These findings also highlight the importance of using complementary approaches when assessing feature relevance, as permutation scores alone may underestimate the value of features that are correlated or interact non-linearly with others.

        \begin{table*}
            \centering
            \caption{Impact of dropping individual bands on photometric redshift performance using kNN and NN models trained on the DxS dataset. Metrics shown are $\sigma_\mathrm{NMAD}$ (normalised median absolute deviation) and MAE. Lower values indicate better performance.}
            \begin{subtable}{0.48\textwidth}
            \centering
            \caption{kNN model}
            \begin{tabular}{lcc}
            \hline
            \textbf{Feature Set} & $\boldsymbol{\sigma_\mathrm{NMAD}}$ & \textbf{MAE} \\
            \hline
            $g$, $r$, $z$, $W1$, $W2$ (baseline) & 0.3090 & 0.2084 \\
            $g$, $r$, $z$, $W2$ (no $W1$)        & 0.3830 & 0.2584 \\
            $g$, $r$, $z$, $W1$ (no $W2$)        & 0.3372 & 0.2274 \\
            $g$, $r$, $z$, $W1$, $W2$, $FUV$, $NUV$ (baseline) & 0.2584 & 0.1743 \\
            $g$, $r$, $z$, $W1$, $W2$, $FUV$ (no $NUV$)        & 0.2580 & 0.1740 \\
            $g$, $r$, $z$, $W1$, $W2$, $NUV$ (no $FUV$)        & 0.2812 & 0.1897 \\
            \hline
            \end{tabular}
            \label{tab:drop_column_results_knn}
            \end{subtable}
            \hfill
            \begin{subtable}{0.48\textwidth}
            \centering
            \caption{NN model}
            \begin{tabular}{lcc}
            \hline
            \textbf{Feature Set} & $\boldsymbol{\sigma_\mathrm{NMAD}}$ & \textbf{MAE} \\
            \hline
            $g$, $r$, $z$, $W1$, $W2$ (baseline) & 0.3171 & 0.2139 \\
            $g$, $r$, $z$, $W2$ (no $W1$)        & 0.3987 & 0.2689 \\
            $g$, $r$, $z$, $W1$ (no $W2$)        & 0.3611 & 0.2435 \\
            $g$, $r$, $z$, $W1$, $W2$, $FUV$, $NUV$ (baseline) & 0.2699 & 0.1821 \\
            $g$, $r$, $z$, $W1$, $W2$, $FUV$ (no $NUV$)        & 0.2839 & 0.1915 \\
            $g$, $r$, $z$, $W1$, $W2$, $NUV$ (no $FUV$)        & 0.2664 & 0.1797 \\
            \hline
            \end{tabular}
            \label{tab:drop_column_results_nn}
            \end{subtable}
            \end{table*}

\vspace*{\fill}

        \section*{Acknowledgements}
            We thank the anonymous reviewers for their helpful comments.

            This research has made use of the NASA/IPAC Extragalactic Database (NED), operated by the Jet Propulsion Laboratory, California Institute of Technology, under contract with the National Aeronautics and Space Administration, and NASA’s Astrophysics Data System Bibliographic Service. Code and data are available upon request.
            
            This work used data from the Sloan Digital Sky Survey (SDSS-IV; \citealt{lyke2020}) and the Dark Energy Spectroscopic Instrument (DESI). SDSS is funded by the Alfred P. Sloan Foundation, the U.S. Department of Energy Office of Science, and the Participating Institutions (\url{http://www.sdss.org}). DESI is managed by Lawrence Berkeley National Laboratory, with support from the U.S. Department of Energy, the National Science Foundation, and international partners (\url{https://www.desi.lbl.gov}). The DESI collaboration acknowledges the privilege of conducting research on I’oligam Du’ag (Kitt Peak), a site of significance to the Tohono O’odham Nation.

\clearpage
\onecolumn
\section{Average performance metrics for 100 runs}
    Tables \ref{tab:DESI_z_evolution} and \ref{tab:DxS_z_evolution} show the averaged performance metrics for 100 runs of both the kNN and NN models for the DESI ($g$, $r$, $z$, $W1$, $W2$) and DxS ($g$, $r$, $z$, $W1$, $W2$, $NUV$, $FUV$) samples.
    
\FloatBarrier
    \begin{center}
    \begin{longtable}{lccccc}
    \caption{Average performance metrics for 100 runs of both the kNN and NN models across varying redshifts, for the DESI ($g$, $r$, z, $W1$, $W2$) sample. Abbreviations are as for Table~\ref{tab:DESI_performance_metrics}.}\\
    \label{tab:DESI_z_evolution}
        \\\hline\\
        \textbf{Name} & \textbf{Corr} & \textbf{EV} & \textbf{$\sigma_\mathrm{NMAD}$} & \textbf{Max err} & \textbf{MAE} \\
        \hline
        \multicolumn{6}{l}{$0.1 < z \leq 1.0$ (\num{11419})} \\ 
        \hline
        kNN & $0.6005 \pm 0.0152$ & $0.3600 \pm 0.0178$ & $0.1342 \pm 0.0031$ & $0.7579 \pm 0.0504$ & $0.0905 \pm 0.0021$ \\
        NN  & $0.5707 \pm 0.0723$ & $0.3269 \pm 0.0792$ & $0.1447 \pm 0.0151$ & $0.6588 \pm 0.0188$ & $0.0976 \pm 0.0102$ \\
        \hline
        \multicolumn{6}{l}{$1.0 < z \leq 2.0$ (\num{45627})} \\ 
        \hline
        kNN & $0.6774 \pm 0.0026$ & $0.4583 \pm 0.0033$ & $0.1883 \pm 0.0023$ & $0.7930 \pm 0.0369$ & $0.1270 \pm 0.0015$ \\
        NN  & $0.7623 \pm 0.0066$ & $0.5808 \pm 0.0102$ & $0.1592 \pm 0.0017$ & $0.7971 \pm 0.0052$ & $0.1074 \pm 0.0012$ \\
        \hline
        \multicolumn{6}{l}{$2.0 < z \leq 3.0$ (\num{25878})} \\ 
        \hline
        kNN & $0.5046 \pm 0.0099$ & $0.2542 \pm 0.0097$ & $0.2389 \pm 0.0047$ & $0.7525 \pm 0.0136$ & $0.1611 \pm 0.0031$ \\
        NN  & $0.6170 \pm 0.0036$ & $0.3797 \pm 0.0043$ & $0.2177 \pm 0.0041$ & $0.7352 \pm 0.0159$ & $0.1469 \pm 0.0027$ \\
        \hline
        \multicolumn{6}{l}{$3.0 < z \leq 6.0$ (\num{4393})} \\ 
        \hline
        kNN & $0.7823 \pm 0.0238$ & $0.6079 \pm 0.0350$ & $0.1687 \pm 0.0040$ & $1.5137 \pm 0.5351$ & $0.1138 \pm 0.0027$ \\
        NN  & $0.9096 \pm 0.0197$ & $0.8263 \pm 0.0349$ & $0.1298 \pm 0.0021$ & $1.1285 \pm 0.1898$ & $0.0875 \pm 0.0014$ \\
        \hline
    \end{longtable}
    \end{center}

    \begin{center}
    \begin{longtable}{lccccc}
    \caption{Average performance metrics for 100 runs of both the kNN and NN models across varying redshifts, for the DxS ($g$, $r$, $z$, $W1$, $W2$, $NUV$, $FUV$) sample. Abbreviations are as for Table~\ref{tab:DESI_performance_metrics}.}
    \label{tab:DxS_z_evolution}
        \\\hline\\
        \textbf{Name} & \textbf{Corr} & \textbf{EV} & \textbf{$\sigma_\mathrm{NMAD}$} & \textbf{Max err} & \textbf{MAE} \\
        \hline
        \multicolumn{6}{l}{$0.1 < z \leq 1.0$ (\num{3540})} \\ 
        \hline
        kNN & $0.5515 \pm 0.0102$ & $0.3035 \pm 0.0121$ & $0.1474 \pm 0.0062$ & $0.5046 \pm 0.0318$ & $0.0994 \pm 0.0042$ \\
        NN  & $0.6993 \pm 0.0079$ & $0.4863 \pm 0.0119$ & $0.1280 \pm 0.0096$ & $0.5073 \pm 0.0744$ & $0.0864 \pm 0.0065$ \\
        \hline
        \multicolumn{6}{l}{$1.0 < z \leq 2.0$ (\num{11754})} \\ 
        \hline
        kNN & $0.7532 \pm 0.0116$ & $0.5641 \pm 0.0176$ & $0.1785 \pm 0.0041$ & $0.6969 \pm 0.0128$ & $0.1204 \pm 0.0028$ \\
        NN  & $0.8427 \pm 0.0420$ & $0.7087 \pm 0.0686$ & $0.1520 \pm 0.0222$ & $0.7475 \pm 0.1031$ & $0.1025 \pm 0.0150$ \\
        \hline
        \multicolumn{6}{l}{$2.0 < z \leq 3.0$ (\num{8001})} \\ 
        \hline
        kNN & $0.6402 \pm 0.0067$ & $0.3945 \pm 0.0154$ & $0.2032 \pm 0.0036$ & $0.7079 \pm 0.0460$ & $0.1371 \pm 0.0024$ \\
        NN  & $0.8474 \pm 0.0076$ & $0.7174 \pm 0.0123$ & $0.1303 \pm 0.0131$ & $0.6151 \pm 0.0316$ & $0.0879 \pm 0.0088$ \\
        \hline
        \multicolumn{6}{l}{$3.0 < z \leq 4.8$ (\num{1293})} \\ 
        \hline
        kNN & $0.8337 \pm 0.0344$ & $0.6955 \pm 0.0577$ & $0.1398 \pm 0.0052$ & $0.8639 \pm 0.0316$ & $0.0943 \pm 0.0035$ \\
        NN  & $0.9360 \pm 0.0042$ & $0.8744 \pm 0.0080$ & $0.1481 \pm 0.0514$ & $0.4280 \pm 0.0734$ & $0.0999 \pm 0.0347$ \\
        \hline
    \end{longtable}
    \end{center}

\clearpage
\onecolumn
\section{Redshift Outlier Predictions}
\begin{figure*}
        \centering
        \includegraphics[width=0.9\linewidth]{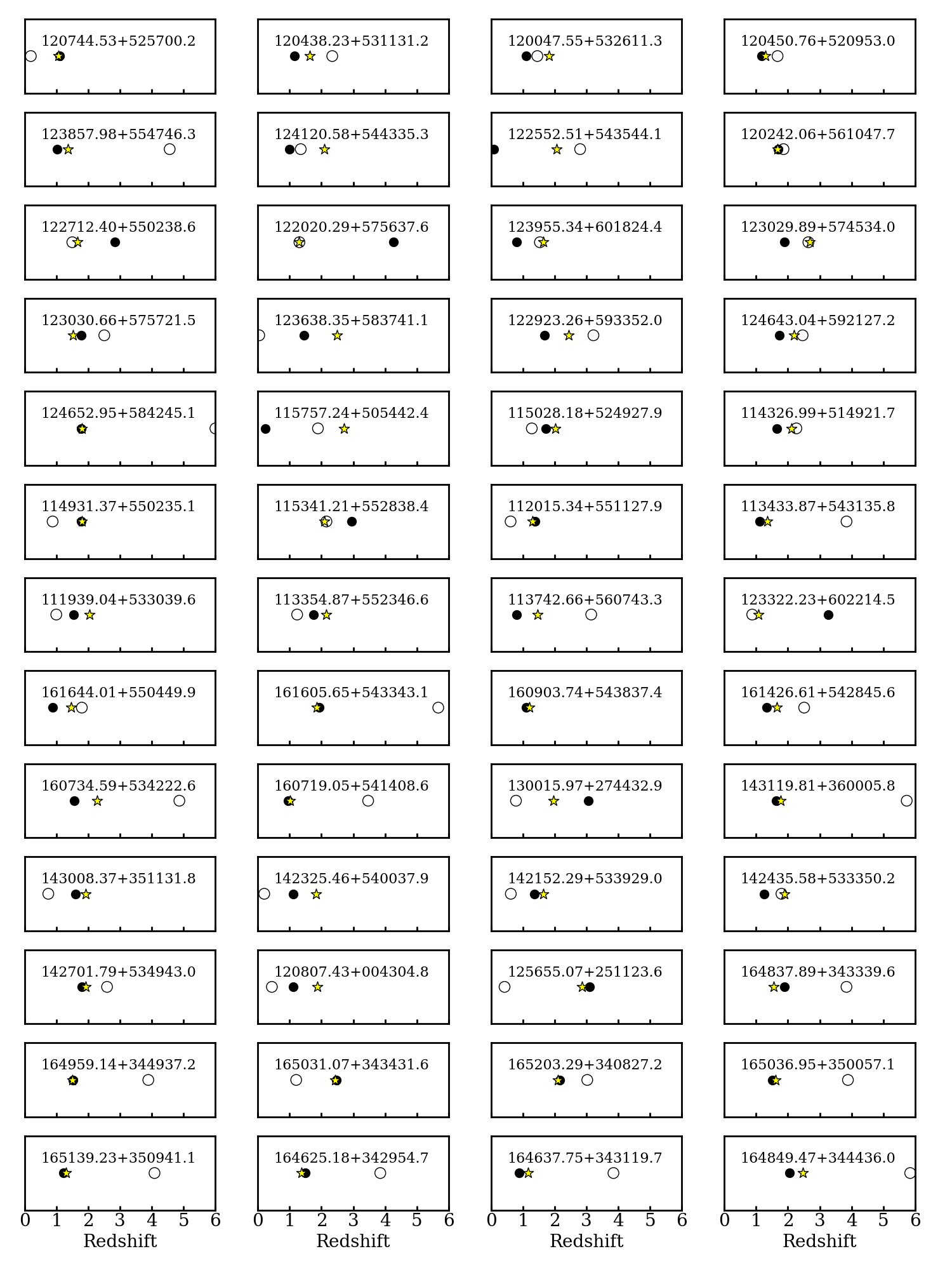}
        \caption{The redshift predictions of a subset of the 107 outliers for which $z_\mathrm{DESI}-z_\mathrm{SDSS} > 0.14$, with 100 runs of the NN trained on the DESI/GALEX ($g$, $r$, $z$, $W1$, $W2$, $NUV$, $FUV$). The filled black markers show the DESI spectroscopic redshift, the unfilled markers the SDSS spectroscopic redshift and the stars the predicted redshift. The label gives the SDSS name of the source. The data is shown in Table~\ref{tab:redshift_outliers}.}
        \label{fig:steve's_figure_17}
    \end{figure*}

    Table \ref{tab:redshift_outliers} displays the SDSS name, spectroscopic redshift from DESI ($z_\mathrm{DESI}$) and from SDSS ($z_\mathrm{SDSS}$) and $z_\mathrm{phot}$ for 100 runs of the kNN trained on the DESI/GALEX fluxes ($g$, $r$, $z$, $W1$, $W2$, $NUV$, $FUV$). A subset of the predictions is visualised in Figure~\ref{fig:steve's_figure_17}.

\begin{longtable}{|l|l|l|l|}
  \caption{Redshift outlier predictions for DESI quasars. 
    A machine-readable version of this table is provided as supplementary material.}
  \label{tab:redshift_outliers}\\
  \hline
  \textbf{SDSS name} & \textbf{$z_\mathrm{DESI}$} 
    & \textbf{$z_\mathrm{SDSS}$} & \textbf{$z_{\mathrm{phot},k\mathrm{NN}}$} \\
  \hline
  \endfirsthead

  \multicolumn{4}{c}%
    {{\bfseries Table~\thetable{} (cont.)}} \\
  \hline
  \textbf{SDSS name} & \textbf{$z_\mathrm{DESI}$} 
    & \textbf{$z_\mathrm{SDSS}$} & \textbf{$z_{\mathrm{phot},k\mathrm{NN}}$} \\
  \hline
  \endhead

  \hline \multicolumn{4}{r}{{Continued on next page}} \\ \hline
  \endfoot

  \hline
  \endlastfoot
        120744.53+525700.2 & 1.1034 & 0.1950 & 1.0708 \\ 
        120438.23+531131.2 & 1.1433 & 2.3400 & 1.6235 \\ 
        120047.55+532611.3 & 1.0957 & 1.4540 & 1.8208 \\ 
        120450.76+520953.0 & 1.1635 & 1.6739 & 1.2968 \\ 
        123857.98+554746.3 & 1.0146 & 4.5623 & 1.3539 \\ 
        124120.58+544335.3 & 1.0003 & 1.3483 & 2.0847 \\ 
        122552.51+543544.1 & 0.0921 & 2.8000 & 2.0524 \\ 
        120242.06+561047.7 & 1.6809 & 1.8600 & 1.6742 \\ 
        122712.40+550238.6 & 2.8389 & 1.5000 & 1.6525 \\ 
        122020.29+575637.6 & 4.2673 & 1.3066 & 1.2893 \\ 
        123955.34+601824.4 & 0.8035 & 1.5350 & 1.6452 \\ 
        123029.89+574534.0 & 1.8808 & 2.6437 & 2.6863 \\ 
        123030.66+575721.5 & 1.7724 & 2.5057 & 1.5230 \\ 
        123638.35+583741.1 & 1.4412 & 0.0424 & 2.4808 \\ 
        122923.26+593352.0 & 1.6855 & 3.2185 & 2.4362 \\ 
        124643.04+592127.2 & 1.7376 & 2.4616 & 2.1906 \\ 
        124652.95+584245.1 & 1.7716 & 6.0009 & 1.7954 \\ 
        115757.24+505442.4 & 0.2260 & 1.8894 & 2.7173 \\ 
        115028.18+524927.9 & 1.7276 & 1.2801 & 2.0112 \\ 
        114326.99+514921.7 & 1.6537 & 2.2660 & 2.1092 \\ 
        114931.37+550235.1 & 1.7873 & 0.8800 & 1.7949 \\ 
        115341.21+552838.4 & 2.9474 & 2.1570 & 2.0980 \\ 
        112015.34+551127.9 & 1.3817 & 0.6130 & 1.3031 \\ 
        113433.87+543135.8 & 1.1057 & 3.8430 & 1.3429 \\ 
        111939.04+533039.6 & 1.5371 & 0.9959 & 2.0352 \\ 
        113354.87+552346.6 & 1.7464 & 1.2350 & 2.1426 \\ 
        113742.66+560743.3 & 0.8008 & 3.1489 & 1.4635 \\ 
        123322.23+602214.5 & 3.2632 & 0.8760 & 1.0656 \\ 
        161644.01+550449.9 & 0.8779 & 1.7993 & 1.4686 \\ 
        161605.65+543343.1 & 1.9264 & 5.6746 & 1.8422 \\ 
        160903.74+543837.4 & 1.1077 & 6.4271 & 1.1936 \\ 
        161426.61+542845.6 & 1.3393 & 2.5097 & 1.6570 \\ 
        160734.59+534222.6 & 1.5619 & 4.8672 & 2.2708 \\ 
        160719.05+541408.6 & 0.9609 & 3.4686 & 1.0211 \\ 
        130015.97+274432.9 & 3.0482 & 0.7820 & 1.9527 \\ 
        143119.81+360005.8 & 1.6312 & 5.7365 & 1.7618 \\ 
        143008.37+351131.8 & 1.5998 & 0.7445 & 1.9146 \\ 
        142325.46+540037.9 & 1.1157 & 0.2010 & 1.8352 \\ 
        142152.29+533929.0 & 1.3702 & 0.6200 & 1.6330 \\ 
        142435.58+533350.2 & 1.2420 & 1.7960 & 1.8991 \\ 
        142701.79+534943.0 & 1.8062 & 2.5924 & 1.9158 \\ 
        120807.43+004304.8 & 1.1022 & 0.4400 & 1.8715 \\ 
        125655.07+251123.6 & 3.1067 & 0.4230 & 2.8520 \\ 
        164837.89+343339.6 & 1.8846 & 3.8413 & 1.5508 \\ 
        164959.14+344937.2 & 1.5142 & 3.8900 & 1.5089 \\ 
        165031.07+343431.6 & 2.4636 & 1.2054 & 2.4384 \\ 
        165203.29+340827.2 & 2.1680 & 3.0243 & 2.1098 \\ 
        165036.95+350057.1 & 1.5093 & 3.8874 & 1.6180 \\ 
        165139.23+350941.1 & 1.2278 & 4.0854 & 1.3028 \\ 
        164625.18+342954.7 & 1.4830 & 3.8503 & 1.3626 \\ 
        164637.75+343119.7 & 0.8761 & 3.8480 & 1.1699 \\ 
        164849.47+344436.0 & 2.0560 & 5.8455 & 2.4611 \\ 
        164756.73+343101.9 & 1.4249 & 5.9711 & 1.6636 \\
        164813.72+345427.3 & 1.8431 & 3.8091 & 1.9107 \\ 
        164903.01+344747.1 & 1.3239 & 3.9994 & 1.6066 \\ 
        164938.61+350118.2 & 0.8684 & 3.8859 & 1.7746 \\ 
        165031.77+352200.2 & 1.2811 & 3.8861 & 1.6640 \\ 
        165121.55+354435.7 & 1.7339 & 3.9147 & 1.7713 \\ 
        165022.10+354245.3 & 1.9101 & 2.7048 & 2.2967 \\ 
        165028.26+355014.0 & 0.9038 & 6.4526 & 1.2311 \\ 
        164534.47+342243.5 & 1.7169 & 2.4125 & 2.4079 \\ 
        164527.62+352655.7 & 1.4801 & 1.2790 & 1.5835 \\ 
        165608.97+350303.4 & 0.9624 & 2.1301 & 1.3526 \\ 
        160832.58+442659.8 & 0.9309 & 1.8467 & 1.8436 \\ 
        153621.26+430637.3 & 0.7747 & 3.0620 & 0.8612 \\ 
        162545.32+414320.5 & 1.9871 & 0.6500 & 1.9211 \\ 
        162118.14+415320.8 & 1.5807 & 1.7788 & 2.0974 \\ 
        162421.09+425243.2 & 0.6574 & 1.9450 & 1.9822 \\ 
        162044.69+434450.4 & 1.7154 & 1.2150 & 2.1543 \\ 
        162427.12+442245.3 & 2.3561 & 3.2375 & 2.4087 \\ 
        153728.60+440246.6 & 0.9349 & 3.4220 & 1.0116 \\ 
        162525.13+441741.5 & 1.5630 & 0.7406 & 1.7337 \\ 
        154918.20+435945.2 & 0.9611 & 0.2055 & 1.0953 \\ 
        162940.01+443919.2 & 1.4396 & 2.1063 & 1.6613 \\ 
        163102.54+442607.2 & 0.8311 & 3.1959 & 1.1509 \\ 
        084347.84+203752.4 & 0.2269 & 5.6085 & 0.1597 \\ 
        110513.65+501216.1 & 1.5794 & 0.7726 & 1.7603 \\ 
        130504.60+334623.5 & 1.9167 & 2.7111 & 2.9692 \\ 
        133704.53+331244.4 & 1.7520 & 2.4910 & 1.9779 \\ 
        130004.92+330411.7 & 1.7953 & 2.5329 & 2.3963 \\ 
        023117.57-003932.9 & 1.0218 & 1.9560 & 1.2245 \\ 
        022836.82+002216.7 & 1.8156 & 2.5767 & 1.8784 \\ 
        115333.54+274404.3 & 1.5767 & 0.7420 & 1.7473 \\ 
        122608.53+305454.2 & 1.7653 & 2.5001 & 1.7710 \\ 
        122155.46+330142.2 & 1.0968 & 3.7896 & 1.2203 \\ 
        122421.52+333857.6 & 1.6043 & 4.8616 & 1.8629 \\ 
        074551.46+162811.6 & 2.4663 & 2.6342 & 3.4707 \\ 
        084510.69+231238.1 & 1.2479 & 1.7852 & 1.3159 \\ 
        093433.41+313920.6 & 0.9982 & 2.1500 & 2.1115 \\ 
        091614.46+310241.8 & 1.0710 & 7.0112 & 1.0101 \\ 
        092347.00+311346.8 & 1.6336 & 2.2290 & 2.2348 \\ 
        091447.69+310934.4 & 2.1578 & 3.0110 & 2.0441 \\ 
        091426.09+315453.6 & 0.9691 & 2.1278 & 1.2812 \\ 
        083252.59+311655.3 & 1.3537 & 1.9017 & 1.6132 \\ 
        100017.13+310410.7 & 0.9185 & 3.3899 & 3.2231 \\ 
        100010.17+314251.2 & 1.9656 & 2.8216 & 3.2097 \\ 
        110324.64+310941.9 & 1.6873 & 2.4552 & 2.1218 \\ 
        110110.21+322937.0 & 1.9887 & 0.1780 & 2.5604 \\ 
        110045.66+325711.7 & 1.1391 & 1.6641 & 1.4297 \\ 
        105617.47+322721.6 & 1.7530 & 2.4306 & 2.0418 \\ 
        095234.80+311839.0 & 1.8527 & 2.6109 & 2.4788 \\ 
        102226.96+322644.8 & 1.1088 & 1.4426 & 1.1764 \\ 
        103527.30+312106.6 & 1.5181 & 0.7037 & 1.6365 \\ 
        095303.80+331326.4 & 0.8653 & 1.7440 & 1.0028 \\ 
        074120.45+334715.7 & 0.4259 & 1.0950 & 1.9823 \\ 
        075651.46+344215.2 & 1.7947 & 5.5047 & 1.9359 \\ 
        160829.51+203701.4 & 2.0910 & 0.6751 & 2.3977 \\
        \end{longtable}
        \clearpage
        \onecolumn

    \clearpage